\documentclass[sigconf,10pt]{acmart}

\usepackage[]{hyperref}

\usepackage{subfigure}
\usepackage{color}
\usepackage{xcolor}
\usepackage{comment}
\usepackage{algorithm,algpseudocode}
\usepackage{rotating}
\usepackage{multirow}
\usepackage{enumitem}
\usepackage{url}
\usepackage{xspace}
\usepackage{subfigure}
\usepackage{amsmath}
\usepackage{tabularx}  
\usepackage{graphicx}
\usepackage{array}
\usepackage{balance}
\usepackage{newtxmath}

\newcommand{\rv}[1]{{\color{black}#1}}

\copyrightyear{2024}
\acmYear{2024}
\setcopyright{rightsretained}
\acmConference[SoCC '24]{ACM Symposium on Cloud Computing}{November 20--22, 2024}{Redmond, WA, USA}
\acmBooktitle{ACM Symposium on Cloud Computing (SoCC '24), November 20--22, 2024, Redmond, WA, USA}
\acmDOI{10.1145/3698038.3698534}
\acmISBN{979-8-4007-1286-9/24/11}

\begin{document}

\newcommand{\name}{\textsc{Deoxys}\xspace}
\newcommand{\model}{\texttt{CauForeDML}\xspace}
\newcommand{\company}{Microsoft\xspace}
\newcommand{\rb}{\textsc{Reboot}\xspace}
\newcommand{\rd}{\textsc{Redeploy}\xspace}
\newcommand{\no}{\textsc{NoOp}\xspace}
\newcommand{\eg}{\emph{e.g.},\xspace}
\newcommand{\ie}{\emph{i.e.},\xspace}
\newcommand{\etal}{\emph{et al.},\xspace}
\newcommand{\etc}{\emph{etc}\xspace}
\newcommand{\saving}{53\%\xspace}
\newcommand{\airsaving}{49.5\%\xspace}
% \DeclareMathAlphabet\mathbfcal{OMS}{cmsy}{b}{n}
%don't want date printed

%make title bold and 14 pt font (Latex default is non-bold, 16 pt)
\title[\name:  A Causal Inference Engine for Unhealthy Node Mitigation]{\name:  A Causal Inference Engine for Unhealthy Node Mitigation in Large-scale Cloud Infrastructure}

\settopmatter{authorsperrow=4}

\author{Chaoyun Zhang}
\affiliation{%
\institution{Microsoft}
  \country{China}
}

\author{Randolph Yao}
\affiliation{%
\institution{Microsoft}
  \country{USA}
}

\author{Si Qin}
\affiliation{%
\institution{Microsoft}
  \country{China}
}

\author{Ze Li}
\affiliation{%
\institution{Microsoft}
  \country{USA}
}

\author{Shekhar Agrawal}
\affiliation{%
\institution{Microsoft}
  \country{USA}
}

\author{Binit R. Mishra}
\affiliation{%
\institution{Microsoft}
  \country{USA}
}

\author{Tri Tran}
\affiliation{%
\institution{Microsoft}
  \country{USA}
}

\author{Minghua Ma}
\affiliation{%
\institution{Microsoft}
  \country{USA}
}

\author{Qingwei Lin}
\affiliation{%
\institution{Microsoft}
  \country{China}
}

\author{Murali Chintalapati}
\affiliation{%
\institution{Microsoft}
  \country{USA}
}

\author{Dongmei Zhang}
\affiliation{%
\institution{Microsoft}
  \country{China}
}

\renewcommand{\shortauthors}{Zhang et al.}

% \setlength{\droptitle}{-3em}
% Use the following at camera-ready time to suppress page numbers.
% Comment it out when you first submit the paper for review.
\thispagestyle{empty}
% \vspace{-1011em}

\begin{abstract}
The presence of unhealthy nodes in cloud infrastructure signals the potential failure of machines, which can significantly impact the availability and reliability of cloud services, resulting in negative customer experiences. Effectively addressing unhealthy node mitigation is therefore vital for sustaining cloud system performance. This paper introduces \name, a causal inference engine tailored to recommending mitigation actions for unhealthy node in cloud systems to minimize virtual machine downtime and interruptions during unhealthy events. It employs double machine learning combined with causal forest to produce precise and reliable mitigation recommendations based solely on limited observational data collected from the historical unhealthy events. To enhance the causal inference model, \name further incorporates a policy fallback mechanism based on model uncertainty and action overriding mechanisms to \emph{(i)} improve the reliability of the system, and \emph{(ii)} strike a good tradeoff between downtime reduction and resource utilization, thereby enhancing the overall system performance.

After deploying \name in a large-scale cloud infrastructure at \company, our observations demonstrate that \name significantly reduces average VM downtime by \saving compared to a legacy policy, while leading to \airsaving lower VM interruption rate. This substantial improvement enhances the reliability and stability of cloud platforms, resulting in a seamless customer experience. 
\end{abstract}


\maketitle

\section{Introduction\label{sec:intro}}
Large-scale cloud infrastructures are composed of millions of computing resources or nodes, each of which is monitored using a heartbeat signal to determine its health status. Occasionally, these signals may be lost due to a variety of reasons such as networking issues, hardware failures, or software bugs, resulting in the node entering an ``unhealthy'' state, which indicates a potential machine failure that can affect all virtual machines (VMs) on that node. The presence of unhealthy events can negatively impact customer experience, as their services and deployments may be disrupted until the machines are mitigated. Despite the existence of proactive systems \cite{levy2020predictive} that predictively mitigate potential node failures before they occur, unhealthy events still frequently occur in large-scale cloud infrastructures \cite{garraghan2014empirical, endo2017minimizing}, which can result in significant cloud outages and loss of millions of dollars \cite{CloudOutages}.

When unhealthy events do happen, the highest priority is to choose an appropriate corrective action to the node, \eg \rb, \rd, for mitigation that minimizes disruption to VMs deployed on that node and accelerates the recovery time \cite{egwutuoha2012proactive}, ensuring a smooth customer experience. However, choosing the most appropriate mitigation action for unhealthy events in a cloud environment is a complex task that presents several challenges. Ideally, immediate action should be taken when an unhealthy event occurs to minimize VM downtime. This however means that only limited information about the event is available, without knowing the root cause that led to the failure. Root cause analysis processes can take several hours or even days \cite{soldani2022anomaly, ma2020diagnosing, liu2024large}, and waiting for too long is not feasible in a cloud environment where prompt response is required. This incomplete information poses significant challenges to the mitigation engine, as it needs to make decisions based on limited signals. Furthermore, root causes of unhealthy events are often heterogeneous \cite{singh2021surviving, chen2023empowering, wang2023root, ding2023tracediag, jiang2023xpert}. Taking different mitigation actions may therefore result in significant variance in the duration of VM downtime \cite{levy2020predictive}. This heterogeneity further increases the reliability requirements of the system, as choosing the wrong mitigation action can potentially cause disruptive impacts to the cloud system.

An effective approach to address these challenges involves the integration of data-driven machine learning techniques \cite{ding2023everything, ufo} into the mitigation engine \cite{colman2016survey}. However, many of these methods rely on online or reinforcement learning \cite{levy2020predictive, wang2022nenya}, necessitating training through interaction with the production environment, which may result in performance regression during the exploration phase. Offline methods are also viable, although they demand a meticulous approach to mitigate potential confounders and ensure the acquisition of reliable experimental data. Otherwise, the predictions may be subject to data bias, resulting in less reliable recommendation \cite{wang2021deconfounded}. A commonly used approach for experimentation to eliminate confounders is \emph{action-level} A/B testing \cite{levy2020predictive}, where different mitigation actions can be compared and evaluated. However, conducting large-scale A/B testing for unhealthy node mitigation may not be feasible in a cloud environment, as it is typically randomized, regardless of the node status. This can lead to potential risks associated with taking incorrect mitigation actions on a large number of nodes, causing system performance regression, which can result in a poor customer experience and impact the reputation of the cloud provider.

This paper introduces \name, an end-to-end causal inference mitigation engine engineered to bridge the aforementioned gap. \name has been seamlessly integrated into \company's expansive cloud computing infrastructure, and it's geared towards furnishing intelligent mitigation strategies in response to the occurrence of unhealthy events. One of its notable features is the capacity to be trained offline exclusively on observational data, obviating the need for action-level A/B testing and online learning. Once trained, it can be promptly deployed to recommend effective mitigation actions upon detecting unhealthy failures, thereby minimizing VM downtime and interruptions. This eliminates the need for the exploratory warm-up stages required by online learning methods, which enhances the reliability and stability of the overarching platform, optimizing the virtual machine experience for customers.

In the inference layer, \name utilizes advanced causal inference techniques \cite{pearl2009causality, zhang2021unified}, including double machine learning \cite{jung2021estimating} and causal forest \cite{wager2018estimation}. These methods allow for the generation of precise and dependable mitigation recommendations using constrained signals from the unhealthy node. Importantly, the models are trained exclusively on observational data, eliminating the necessity for expensive and disruptive randomized action-level experiments. In addition, \name is capable of extracting simple logical ``if-else'' like policies from the data, which highlight the most important features driving its decision-making process. This interpretability is crucial in a cloud system for accountability, debugging, and troubleshooting purposes \cite{chakraborttii2020improving}, allowing system operators and administrators to gain insights into the decision-making process of \name.

Furthermore, \name incorporates several dedicated system design elements to ensure its robustness and resource requirements are met within a cloud computing environment \cite{agarwal2023unlocking}. One such design feature is the inclusion of policy fallback, which enables \name to automatically adjust its decision when it has low confidence. This approach helps to minimize the risk of erroneous decisions by the model, thereby enhancing its system reliability. Furthermore, \name is designed to override an action if it determines that the gain in downtime reduction is minor compared to another action, but the tradeoff is a potential node resource saving. This allows \name to strike a balance between minimizing downtime and optimizing capacity utilization. Additionally, if unhealthy events occur repeatedly at the same node within a short time period, another action override will be triggered to revert the \rb to \rd, which prevents repeated failures from jeopardizing the system. All of these designs ensure that the overall system performance is optimized while maintaining robustness and capacity requirements.

We deployed \name in a large-scale cloud infrastructure at \company. Our findings revealed that \name was able to effectively reduce the average VM downtime compared to the legacy policy for unhealthy nodes, while leading to substantially lower VM interruption rate. This has positioned it as a critical component in the mitigation strategy of the cloud platform at \company. Overall, our contributions are summarized as follows:
\begin{itemize}[leftmargin=*]
    \item  We propose \name, a causal inference based unhealthy failure mitigation engine that recommends mitigation actions, without the need for large-scale, interruptive \emph{action-level} A/B testing for data collection, and the exploratory warm-up stages required by online learning methods.
    \item  We design a policy fallback and action override mechanisms in \name to improve the reliability of the system, while achieving a balance between downtime reduction and resource utilization.
    \item Offline comparisons using a high-fidelity simulator show that \name can achieve over 14\% reduction in downtime compared to the state-of-the-art mitigation policy.
    \item We deploy \name as an unhealthy mitigation engine in a large-scale cloud infrastructure at \company. Our observations show \saving downtime reduction, as well as \airsaving lower VM interruption rate compared to a legacy policy. These reductions potentially lead to better customer experience and huge business value.
    \item  We extract logical policies from \name to gain insights from data and understand the features that drive its predictions, unveiling the decision-making process of \name.
\end{itemize}
To the best of our knowledge, \name represents the first deployment of an unhealthy mitigation engine based on causal inference in a large-scale cloud infrastructure.

% \vspace*{-0.5em}
\section{Background and Overview}
In a large-scale cloud system, a node is marked as ``unhealthy'' if a central controller loses communication with it for a specified time threshold. In such cases, the controller sends a diagnostic request to retrieve the status of the affected node. If the node is unable to self-recover, the controller must take mitigation action to restore the VMs operating on the node.

% \vspace*{-0.5em}
\subsection{System \& Objective}
The aim of this research is to address the issue of smart failure mitigation in cloud systems when aforementioned unhealthy events occur. Specifically, we focus on the VM host environment, which represents a node in a cloud computing platform. This environment is composed of a complex stack of components, including guest OSes, guest agents, hypervisor, host OS, host agents, firmware, and hardware. The node is connected to various compute services, referred to as the controller, responsible for resource provisioning and management actions such as creating and destroying VMs.

The primary objective of this study seeks to develop advanced and intelligent policies that automatically identify the most effective mitigation action to minimize VM downtime and interruption when an unhealthy node is detected, so as to improve the reliability and availability of cloud systems \cite{nabi2016availability, vishwanath2010characterizing}.

\begin{figure}[t]
\centering
\includegraphics[width=\columnwidth]{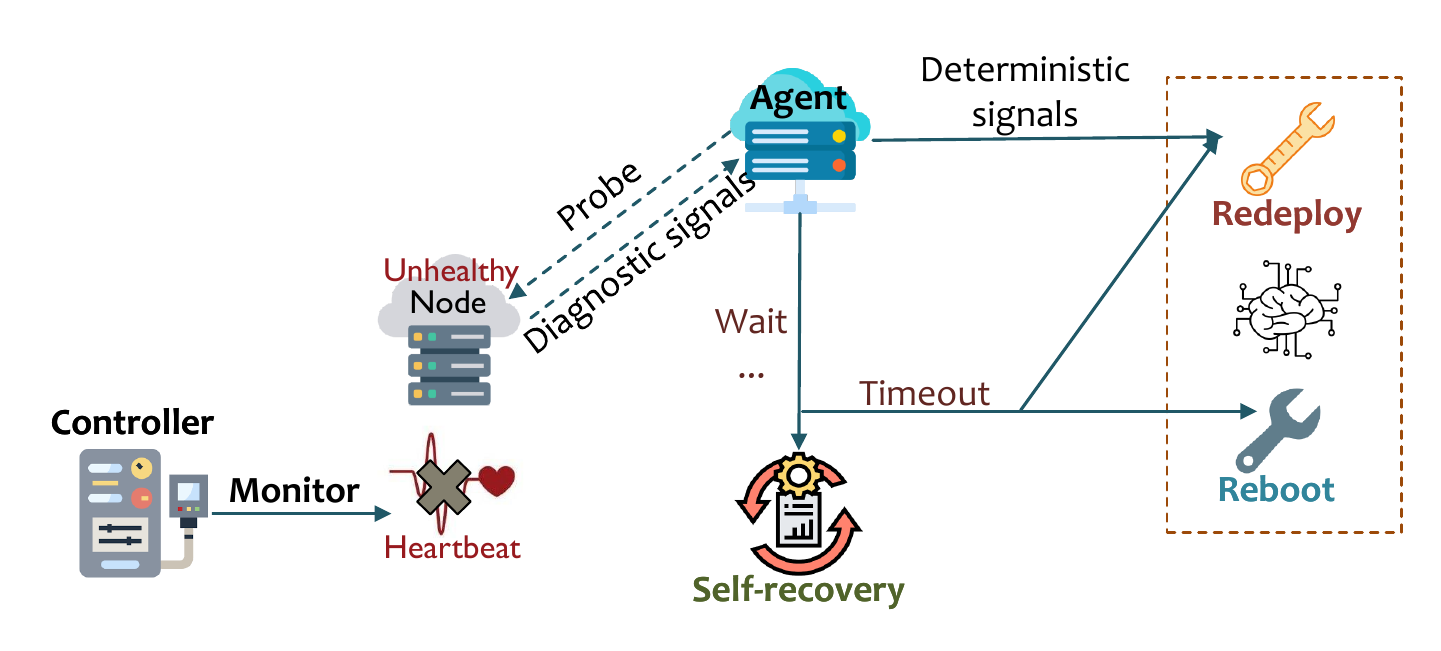}
\vspace*{-2.em}
\caption{The workflows of unhealthy events and mitigation.
\label{fig:unhealthy}}
% \vspace*{-2.5em}
\end{figure}

% \vspace*{-0.5em}
\subsection{Unhealthy Event Workflow\label{sec:workflow}}
In the context of production settings, modern cloud computing platform usually employs a controller to oversee and regulate the operational status of nodes. Nodes, which are the fundamental computational units in the cloud infrastructure, can exist in a variety of different states, the most common of which is the ``ready'' state wherein they are fully functional. On occasion, a node may transit into an ``unhealthy'' state, indicating its inability to respond to requests from the controller within a predetermined timeframe. This state may arise due to various factors, such as software and hardware malfunctions or network issues. In such scenarios, the controller may intervene to mitigate the unhealthy node by taking different actions, such as waiting for the node to self-recover or executing alternative mitigation measures. Nevertheless, such mitigations incur downtime for the node.

The specific set of procedures for managing unhealthy nodes are shown in Fig.~\ref{fig:unhealthy}. Firstly, the central controller loses communication with the heartbeat sender of a node, and if this communication loss persists for a specified duration, the node state is transitioned to unhealthy. Subsequently, if the node state remains unhealthy for a further period, the controller sends a signal to an agent, which triggers the unhealthy workflow in the system. The agent then performs several node diagnostic procedures, to further collect node data understand its unhealthy state. If the diagnostic results yield deterministic signals, such as an uncorrectable error, the agent immediately takes the node out of rotation and \rd the customer workload to a different healthy node. Alternatively, the agent waits for an unhealthy timeout, before taking any invasive action. 
Finally, if the node remains unhealthy after the timeout, the agent attempts to mitigate the situation by performing an in-place repair, such as a node \rb, or triggering a \rd by transiting all VMs to other healthy nodes. Choosing an appropriate mitigation action is critical because it can lead to different VMs downtime. This forms the foundation objective of this research.

\rv{We note that, grey failures, where a node experiences performance degradation but does not fully fail, typically do not trigger the unhealthy node mitigation process because the node's heartbeat remains active. Therefore, such failures are handled by mechanisms outside of \name.}

% \vspace*{-0.5em}
\subsection{Diagnostic Signals\label{sec:signal}}

The proposed \name and the controller rely on diagnostic signals to probe the system status of unhealthy nodes. These signals provide important insights that can reduce the observability gap \cite{huang2018capturing} and facilitate effective decision-making.

\begin{itemize}[leftmargin=*]
    \item \textbf{VM Information}: Number of VMs, types of VMs, session types, and whether important workloads are running on the node. These are used to understand the current workload and resource utilization of the node to decide on an appropriate mitigation action.
    \item \textbf{Network Information}: Network connectivity status of the node, used to determine if network issues are contributing to the unhealthy node status to make appropriate mitigation decisions.
    \item \textbf{Error Code}: Helps diagnose the root cause of the issue. Though sometimes missing, it can provide valuable diagnostic information in certain scenarios for more effective mitigation.
    \item \textbf{Unhealthy Repeat Times Signal}: Frequency of the node becoming unhealthy in the past. It is an indicator of impending failure, which helps in proactive mitigation actions.
     \item \textbf{Uncorrectable Error Tag Signal}: Indicates whether the node can be corrected. It is to decide whether to take corrective actions or to leave the node for human investigation.
\end{itemize}

Overall, these diagnostic signals provide a rich set of features that help understand the state of the unhealthy node and inform appropriate mitigation decisions.

% \vspace*{-0.5em}
\subsection{Mitigation Actions \& Legacy Policy\label{sec:policy}}
In the event that a node becomes unhealthy, it is incumbent upon the central controller to take one of two actions to restore the VMs operating on the affected node. Unless the node undergoes a self-recovery, the central controller is responsible for ensuring the proper functioning of the affected VMs. These actions include:
\begin{enumerate}[leftmargin=*]
    \item \textbf{\rb} action constitutes a mechanism that facilitates the preservation of VM state in the event of a host Operating System (OS) reboot \cite{candea2004microreboot}. Fundamentally, the process entails reloading the host OS kernel into memory while maintaining the persistence of the VM and device state to the reloaded kernel. The reboot is then executed with the loaded kernel, and the persisted state is retained, while the rest of the state in the previous kernel is discarded. Upon the commencement of the reloaded kernel, the preserved state is restored, albeit with a momentary interruption. In \rb, VMs are restored at the same node without requiring additional resources.
    \item \textbf{\rd} action is an approach that ensures the seamless transfer of a running VM from one host to another using a live migration \cite{he2023taxonomy} or service healing \cite{gill2019radar} engine, with minimal disruptions. The process involves migrating the VM's memory, processor, and virtual device state. The engine facilitates iterative copying of the VM's memory pages while preserving a dirty page set for the VM on the source host. 
    After the VM has been halted, the engine synchronizes the dirty state with the target host and resumes the VM on the target host. Notably, various reasons may lead to the failure of \rd, \eg shortage of node resource. In the case of \rd, all VMs will encounter a brief pause similar to a \rb. Note that the \rd action entails the migration of VMs to a different node, which requires additional resources.
\end{enumerate}
The current study is centered on the scenario whereby the controller loses the heartbeat for a specified time interval, prompting the need to select either the \rb or \rd as a mitigation action. 

\noindent \textbf{Legacy Policy} The management of unhealthy nodes previously relies on a legacy mitigation policy that employs heuristic rules for making mitigation decisions based on diagnostic signals. It predicates on an offline data collection approach centered on the success probability of the repair action. It entails waiting for a predefined, heuristic duration derived from data accumulated over a two-year period preceding the policy modification. Should the machine fail to spontaneously recover within this time frame, a \rb was initiated, followed by an extended wait before determining whether to \rd the VMs to an alternate machine. Notably, the heuristic waiting period was solely determined by the hardware type and limited diagnostic signals, which does not take into account any of the workload characteristics or other environmental attributes. 
This simple policy may overlook important factors and criteria between mitigation action and VM downtime. In this study, we compare the performance of \name with this heuristic legacy policy, as it is the only policy deployed in production.

% \vspace*{-0.5em}
\subsection{Key Performance Indicators\label{sec:kpi}}

% \mh{I suggest separating this subsection into a new section and first presenting the end-to-end impact}
The key performance indicators (KPIs) of the unhealthy mitigation task are mainly evaluated from two perspectives: \emph{(i)} Average VM Downtime (AVD); and \emph{(ii)} Annual Interruption Rate (AIR) \cite{levy2020predictive}. For a single unhealthy mitigation request on a machine node $\mathcal{N}$, AVD is defined as the average downtime by all VMs on that node:
\begin{align}
\mathbf{AVD} = \frac{\sum_{\mathrm{VM} \in \mathcal{N}} Y_{\mathrm{VM}}}{\mathrm{VM\ count}}.
\end{align}
This metric represents the primary objective of \name and is the key optimized metric for our system. Additionally, we also consider the AIR, which is defined as:
\begin{equation}
    \mathbf{AIR} = \frac{I_{\mathcal{T}}}{L_{\mathcal{T}}} \times 365 \times 100.
\end{equation}

\rv{
In this equation, $\mathcal{T}$ represents the duration of the measurement interval, expressed in days. For the purposes of this paper, we set $\mathcal{T}$ to 4 months, corresponding to the full experimental period. The variable $ I_{\mathcal{T}}$ denotes the number of VM interruptions that occurred within the interval $\mathcal{T}$, while $L_{\mathcal{T}}$ represents the total VM-lifetime within the same period, measured in days. The constant 365 is used to scale the rate to an annualized figure, and the factor of 100 adjusts the result to represent the number of interruptions per 100 VMs annually.
}

\rv{
In the context of this study, a VM interruption refers to a forced reset of a VM, initiated by automated mitigation actions. The AIR metric specifically captures the frequency of these unhealthy events per 100 VMs in a year and is independent of their duration. A lower AIR value corresponds to fewer interruptions. We choose this metric for several reasons. First, even brief interruptions can severely degrade user experience, particularly for latency-sensitive applications such as online gaming. Additionally, based on feedback from our customers, frequent interruptions are perceived as more disruptive and frustrating than a single, longer period of downtime. AIR is thus a meaningful measure of service reliability and user experience in cloud environments, as it effectively quantifies the frequency of such interruptions.
}

Note that due to the dynamic relationship between the mitigation policy and the node state, the mitigation actions can significantly impact the AIR. Inappropriate actions may lead to new unhealthy events and VM interruptions occurring in quick succession. Although the AIR is not a direct optimized objective of \name, we closely monitor this metric to ensure that the policy derived from \name does not result in significant regressions on AIR.

\subsection{Challenges\label{sec:challenge}}
In addressing the unhealthy event mitigation problem, we acknowledge several challenges that need to be overcome.
\begin{itemize}[leftmargin=*]
    \item \textbf{Challenge 1: Observational Data Contamination}: The data available for model training is observational, as the legacy policy, on which the mitigation engine relies, does not conduct randomized experiments for each action. Observational data are contaminated by confounding variables, making it challenging to discern the true causation between mitigation action and downtime.
    \item \textbf{Challenge 2: Incomplete Information}: The lack of complete information regarding the unhealthy node and its root causes poses a significant challenge, particularly when utilizing data-driven methods. The uncertainty stemming from incomplete information must be considered in the mitigation design.
    \item \textbf{Challenge 3: Resource Limitations}: The \rd action, which requires migrating all VMs from an unhealthy node to a healthy one, may be limited by the system's available resources. This poses a practical challenge in ensuring that the system can perform the \rd action without overwhelming the system's resources or causing additional disruptions.
    \item \textbf{Challenge 4: Mitigation Engine Mistakes}: The potential for mistakes by the mitigation engine is a significant concern, as a single wrong decision could potentially lead to disastrous consequences. It is crucial to incorporate a protection mechanism to minimize the risk of erroneous decisions by the engine.
\end{itemize}
Addressing these challenges in the design of \name is essential to ensure the reliability of the system.

% \vspace*{-.5em}
\section{Causal Modeling \label{sec:causal}}
The stringent requirements for high reliability and interpretability in unhealthy mitigation render causal inference highly suitable, a perspective often overlooked by traditional systems. Specifically, for causal models: \emph{(i)} they learn an unbiased policy from observational data generated by the legacy policy, obviating the need for costly action-level A/B testing for data collection;  \emph{(ii)} they can be trained offline, without the need of warm-up stage on online environment; \emph{(iii)} they provide high explainability by extracting if-else-like policies (see Sec.~\ref{sec:interpreter}). \name thus leverages causal inference as the foundation for constructing its mitigation policy.

To first investigate the relationship between node downtime and mitigation actions, we construct a causal model to explicitly analyze unhealthy events for optimizing the mitigation decision-making process. This process involves:
\begin{enumerate}[leftmargin=*]
    \item Constructing a causal graph that delineates the interrelationships between various system variables.
    \item Formulating a causal machine learning model based on the causal graph, which serves to rectify data biases inherent in legacy policies and estimate the causal effect of mitigation actions on VM downtime.
    \item Predicting the potential downtime difference when a mitigation action is selected using the causal model. We then base our mitigation policy on the predicted values.
\end{enumerate}
We detail these processes in the following subsections.

\begin{figure}[t]
\centering
\includegraphics[width=\columnwidth]{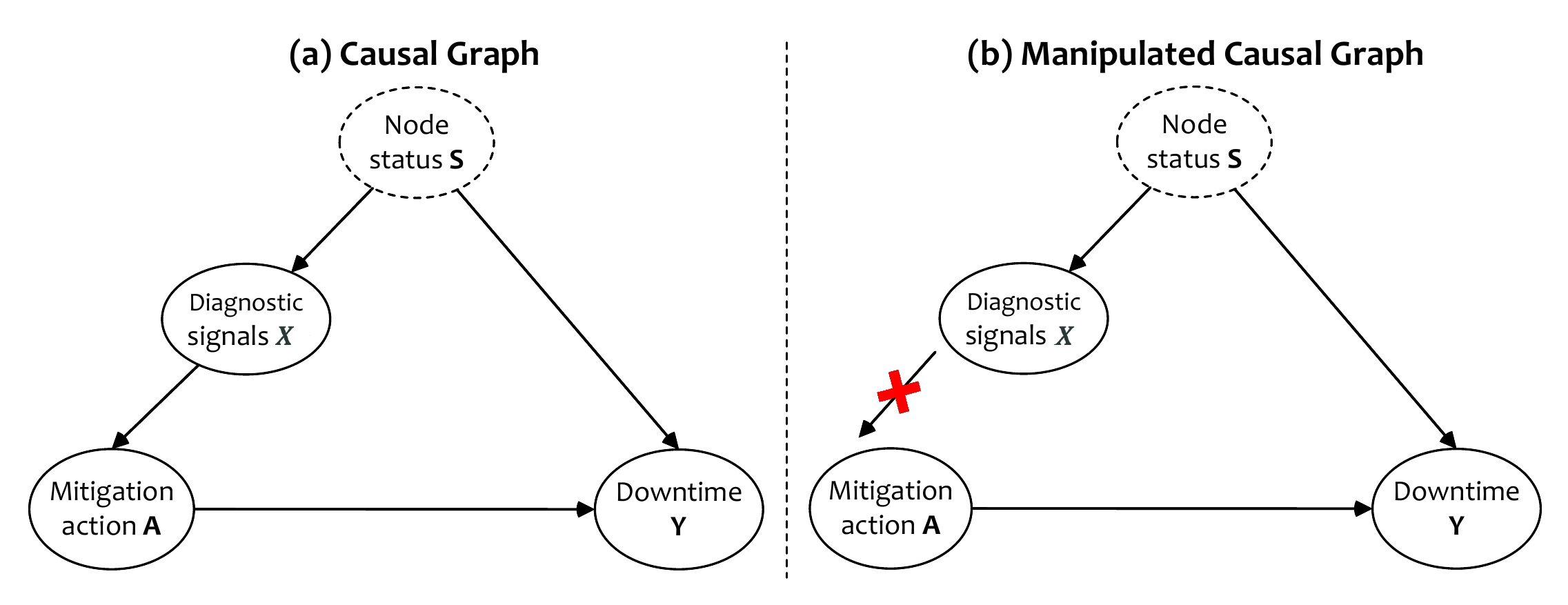}
\vspace*{-2.4em}
\caption{The causal graph (a) and the manipulated causal graph (b) of the unhealthy node mitigation.
\label{fig:causalgraph}}
% \vspace*{-1.3em}
\end{figure}

% \vspace*{-1.em}
\subsection{A Causal View of Unhealthy Events}
Based on the unhealthy node workflow shown in Fig.~\ref{fig:unhealthy}, we can derive an abstract graphical causal model that represents the causal relationships depicted in the left plot of Fig.~\ref{fig:causalgraph}(a). 

The causal graph constructed comprises four sets of variables, namely $S$, $\mathbf{X}$, $A$, and $Y$:
\begin{itemize}[leftmargin=*]
    \item $S$ captures the overall status of an unhealthy node, such as the software and hardware conditions, and the root cause of the issue \etc. 
    % Armed with this information, we are able to take decisive remedial measures. 
    However, the complete status $S$ is not fully observable. $S$ can be partially characterized by the observable diagnostic signals $\mathbf{X}$.
    \item $\mathbf{X}$ comprises $N$ diagnostic signals (described in Sec.~\ref{sec:signal}) that are obtained by probing the status of the unhealthy node, \ie $\mathbf{X} = \{X_1, X_2, \cdots, X_N\}$. These signals are fully observable, enabling us to design appropriate mitigation strategies. Nonetheless, $\mathbf{X}$ provides less information compared to $S$, thereby leaving a greater degree of uncertainty. 
    \item $A$ denotes the mitigation action taken after an unhealthy event, where $A \in \{0, 1\}$. The value $A=0$ corresponds to the \rb, while $A=1$ denotes the \rd.
    \item $Y$ quantifies the duration of VM downtime that occurs subsequent to the mitigation action. It is calculated as the average value of downtime across all VMs in a node.
\end{itemize}
In addition, edges in the causal graph depict the causal relationships between the variables. Specifically,
\begin{itemize}[leftmargin=*]
    \item $S\to \mathbf{X}$: The status of the node $S$ influences the set of diagnostic signals $\mathbf{X}$. Given $S$ is not fully observable, $\mathbf{X}$ becomes the only signals that can be used to profile the node state and formulate a proper mitigation policy.
    \item $\mathbf{X} \to A$: In the legacy policy, the mitigation action $A$ is determined by the diagnostic signals $\mathbf{X}$, based on heuristic rules and subject to randomness, as described in Sec.~\ref{sec:policy}.
    \item $(S, A) \to Y$: The duration of VM downtime $Y$ is influenced by both the node status $S$ and the mitigation action $A$. This is because the recovery of VMs follows different workflows for different mitigation actions, each of which is associated with a different processing time. 
\end{itemize}

The graph depicts that variables $S$ and $\mathbf{X}$ act as confounding factors \cite{pearl2009causality}, as they simultaneously influence both the mitigation action $A$ and the downtime $Y$. This arises due to the limitations of collecting only observational data from the legacy policy.

\textbf{Problem:} Confounding factors have the potential to distort the relationship of $P(Y|A)$ from the true causation between mitigation action ($A$) and VM downtime ($Y$), leading to imprecise and biased estimates of their true effect. This can compromise decision-making, as illustrated by Simpson's paradox \cite{wagner1982simpson}, where these factors can make interventions appear more or less effective than they truly are by influencing both the mitigation action and the VM downtime. 
To prove, we transform the correlation probability $P(Y|A)$ as follows:
\begin{align}\label{eq:correlations}
     P(Y|A) &\stackrel{(1)}{=} \sum_{\mathbf{X}} P(Y, \mathbf{X}|A) \nonumber\\
                &\stackrel{(2)}{=} \sum_{\mathbf{X}} P(Y|A,  \mathbf{X})P(\mathbf{X}|A) \nonumber\\
                &\stackrel{(3)}{\propto} \sum_{\mathbf{X}} P(Y|A,  \mathbf{X})P(A|\mathbf{X})P(\mathbf{X})
\end{align}
The concept of marginal distribution underpins Line (1). The subsequent Lines (2) and (3) are derived from the Bayesian rule. In particular, the term $P(Y|A, \mathbf{X})P(A|\mathbf{X})$ in Line (3) is influenced by the observed data. Specifically, the selection of a particular action for a given $\mathbf{X}$ in the dataset $P(A|\mathbf{X})P(\mathbf{X})$ leads to an increase in the value of $P(Y|A)$ and consequently introduces bias in the estimation.

While one remedy approach can be \emph{action-level} A/B testing, where actions are taken independently of diagnostic signals $\mathbf{X}$, this is impractical as it is  prohibitively expensive and could disrupt the system. Instead, causal inference provides a method to deconfound the mitigation policy without the need for action-level A/B testing,  ensuring more accurate assessments of mitigation effectiveness and addressing the observational data contamination in \textbf{Challenge 1}.

% \vspace*{-0.5em}
\subsection{Deconfounding the Mitigation Policy\label{sec:backdoor}}
To accurately estimate the causal effect of a mitigation action ($A$) on VM downtime ($Y$), we need to account for confounding factors. Our goal is to calculate $P(Y|do(A))$ instead of $P(Y|A)$. $P(Y|do(A))$ represents the causal effect of actively intervening on $A$ to see its direct impact on $Y$ and helps us overcome biases introduced by confounders that could distort the observed relationships.

\subsubsection{$P(Y|do(A))$ vs. $P(Y|A)$}
The probability distribution $P(Y|do(A))$ differs from the $P(Y|A)$, as follows:
\begin{itemize}[leftmargin=*]
    \item \textbf{$P(Y|do(A))$}: This represents the scenario where we actively manipulate $A$ to see its direct effect on $Y$. It isolates the effect of $A$ on $Y$ by removing the influence of confounders $\mathbf{X}$.
    \item \textbf{$P(Y|A)$}: This represents the observed association between $A$ and $Y$, which can be affected by confounders and does not imply causation.
\end{itemize}
To estimate $P(Y|do(A))$, we need to block the path from the confounders ($\mathbf{X}$) to $A$ and reconstruct the causal graph as shown in Fig.~\ref{fig:causalgraph} (b), to ensure that the relationship between $A$ and $Y$ is purely causal. This can be achieved by backdoor adjustment.

\subsubsection{Backdoor Adjustment}
We use the backdoor adjustment \cite{wang2021deconfounded, chen2023bias, he2023addressing} technique to estimate $P(Y|do(A))$ using observational data. According to the backdoor criterion \cite{maathuis2015generalized}:
\begin{itemize}[leftmargin=*]
    \item \textbf{Blocking Backdoor Paths:} $\mathbf{X}$ blocks all backdoor paths from $A$ to $Y$.
    \item \textbf{No Descendants:} $\mathbf{X}$ does not include any descendants of $A$.
\end{itemize}
With these criteria met in Fig.~\ref{fig:causalgraph}, we can translate the causal estimand $P(Y|do(A))$ into a statistical estimand that can be calculated using observational data:
\begin{align}\label{eq:backdoor}
     P(Y|do(A)) &= \sum_{\mathbf{X}} P(Y|A, \mathbf{X})P(\mathbf{X}).
                % &= \sum_{X_n \in \mathcal{X}_n}\sum_{n=1}^N P(Y|A, X_n)\prod_{n=1}^N P(X_n).
\end{align}
Note that the unobservable variable $S$ is eliminated from the equation after the adjustment, as blocking $\mathbf{X} \to A$ is sufficient to satisfy the backdoor criterion. This equation allows us to emulate aspects of experimental design within observational studies, providing a more precise estimation of the causal effect between $A$ and $Y$, even in the presence of confounding variables. We apply the backdoor adjustment technique to our dataset to estimate the Individual Treatment Effect (ITE). This helps us predict the difference in downtime when implementing one mitigation action over another, informing the design of our policy.

% \vspace*{-0.5em}
\subsection{Individual Treatment Effect Estimation}

Using the backdoor adjustment equation, we can train machine learning models to calculate the Individual Treatment Effect (ITE) \cite{shalit2017estimating, zhang2021unified} for each unhealthy event. The ITE, denoted as $\tau_i$, is defined as $\tau_i := P(Y_i|do(A_i=1)) - P(Y_i|do(A_i=0))$. This represents the potential downtime difference when choosing one mitigation action over another:
\begin{itemize}[leftmargin=*]
    \item \textbf{Positive $\tau_i$ ($\tau_i > 0$):} Indicates that choosing action $A=1$ (e.g., \rd) results in longer downtime for event $i$, so we should choose action $A=0$ (e.g., \rb).
    \item\textbf{ Negative $\tau_i$ ($\tau_i < 0$):} Indicates that choosing action $A=0$ (e.g., \rb) results in longer downtime for event $i$, so we should choose action $A=1$ (e.g., \rd).
\end{itemize}
By selecting the optimal mitigation action based on the sign and value of $\tau_i$, we can design an effective mitigation policy to minimize overall VM downtime.

Let $Y_i(0)$ and $Y_i(1)$ represent the potential outcomes of VM downtime when taking actions $A=0$ and $A=1$ for unhealthy request $i$, \ie $P(Y_i|do(A_i=1))$ and $P(Y_i|do(A_i=0))$ respectively. Using the backdoor adjustment, the ITE for request $i$ is:
\begin{align}\label{eq:ite}
    \tau_i  :&= E[Y_i(1) | \mathbf{X}_{i}] - E[Y_i(0) | \mathbf{X}_{i}], \nonumber\\
            :&= \sum_{\mathbf{X}_{i}} E[Y_i(1) | A_i = 1, \mathbf{X}_{i}]P(\mathbf{X}_{i}) \nonumber\\
            &- \sum_{\mathbf{X}_{i}} E[Y_i(0) | A_i = 1, \mathbf{X}_{i}]P(\mathbf{X}_{i}), \nonumber\\
            :&=  E[Y_i(1) | A_i = 1, \mathbf{X}_{i}] -  E[Y_i(0) | A_i = 1, \mathbf{X}_{i}]. 
\end{align}
By estimating the downtime difference for each individual unhealthy request, we can optimize the mitigation policy to minimize the downtime duration for each event, consequently reducing the overall VM downtime for the entire system.

\subsection{Double Machine Learning\label{sec:dml}}
\begin{figure}[t]
\centering
% \vspace*{-1.em}
\includegraphics[width=1\columnwidth]{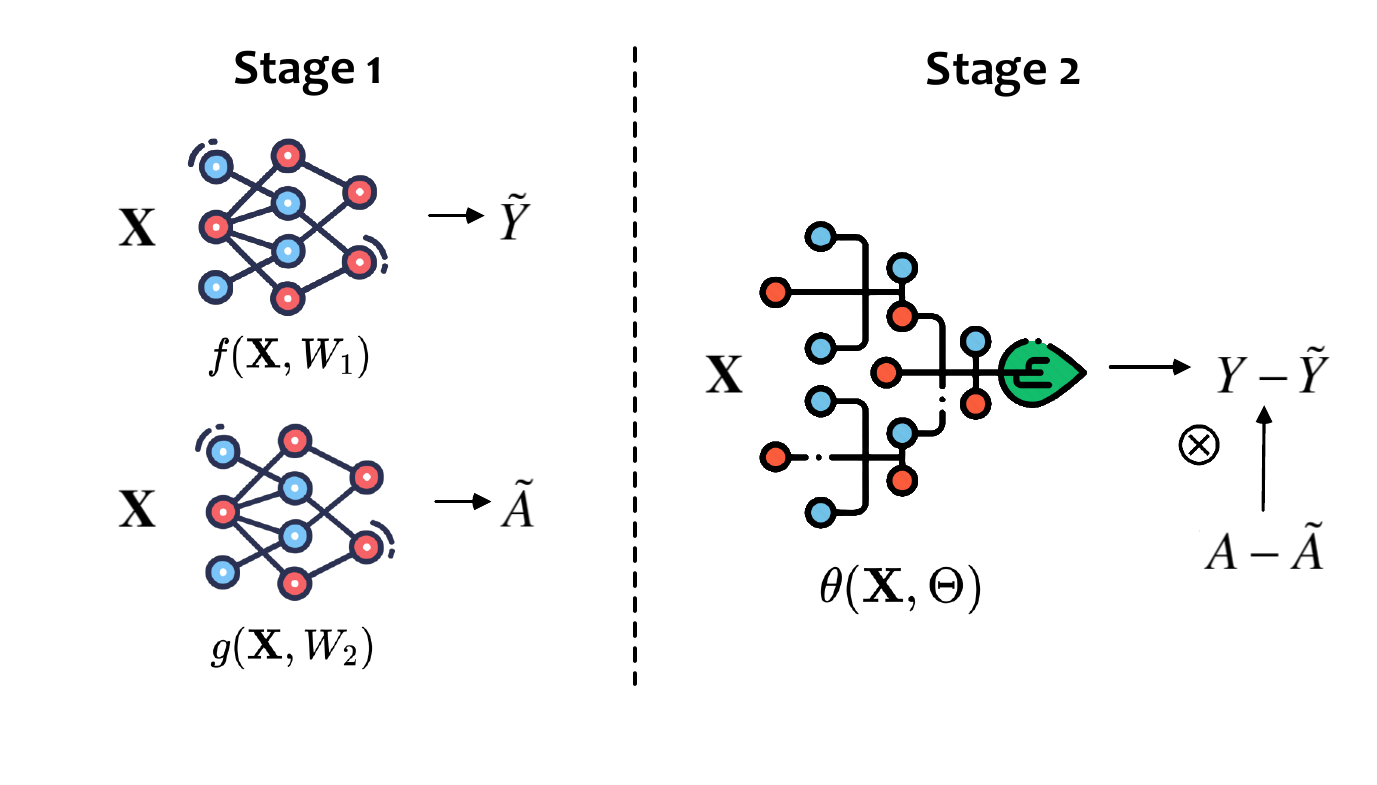}
\vspace*{-2.5em}
\caption{The two-stage DML framework.
\label{fig:dml}}
% \vspace*{-1.em}
\end{figure}

We use Double Machine Learning (DML) to design our mitigation policy for unhealthy requests. DML is a powerful method for estimating ITEs \cite{chernozhukov2018double, jung2021estimating, lewis2021double, lewis2020double}, especially when dealing with high-dimensional data where traditional statistical methods fall short. In our context, we have a large number of diagnostic signals ($\mathbf{X}$) and a complex relationship between the mitigation action ($A$) and the VM downtime ($Y$). DML helps us accurately estimate the causal effect of $A$ on $Y$ by controlling for these confounders. While DML does not directly implement the backdoor adjustment as traditionally defined in causal inference, it aligns with the principles of backdoor adjustment by aiming to control for confounders to accurately estimate causal effects. DML operates in two main stages, namely \emph{(i)} prediction stage and \emph{(ii)} residuals stage, as shown in Fig.~\ref{fig:dml}.

\textbf{Prediction Stage:} We first build two models: One to predict the VM downtime ($Y$) using the diagnostic signals ($\mathbf{X}$). Another to predict the mitigation action ($A$) using the same diagnostic signals.
Mathematically, we represent these models as:
\begin{align}\label{eq:dml_1}
\tilde{Y} &= f(\mathbf{X}, W_1), \quad \tilde{A} = g(\mathbf{X}, W_2).
% \tilde{A} &= g(\mathbf{X}, W_2).
\end{align}
Here, $f$ and $g$ are our predictive models, parameterized by $W_1$ and $W_2$ respectively. $\tilde{Y}$ and $\tilde{A}$ are the predicted values.

\textbf{Residuals Stage:} Next, we use the predictions from the first stage to perform a residuals-on-residuals regression. This isolates the effect of the mitigation action ($A$) on the VM downtime ($Y$):
\begin{align}\label{eq:dml_2}
Y - \tilde{Y} = \theta(\mathbf{X}, \Theta)\cdot (A - \tilde{A}) + \epsilon.
\end{align}
Here, $\theta$ is our final model, parameterized by $\Theta$, and $\epsilon$ is a noise term. We optimize $\Theta$ to minimize the error in this regression:
\begin{align}\label{eq:loss}
\tilde{\Theta} = \arg \min_\Theta \mathbb{E} \left [ \left((Y - \tilde{Y}) - \theta(\mathbf{X}, \Theta)\cdot (A - \tilde{A})\right)^2  \right ].
\end{align}

The key idea behind DML is to remove the bias introduced by confounders. By predicting $Y$ and $A$ first and then focusing on the residuals, we effectively isolate the true impact of $A$ on $Y$. This two-stage process ensures that our estimates are not skewed by other factors \cite{neyman1959optimal}. To further reduce bias, DML uses a technique called cross-fitting. This involves splitting the data into different subsets and training the models on these subsets separately. Think of it as getting multiple opinions before making a decision, which helps average out biases and leads to more reliable estimates.

Once trained, our final model $\theta(\mathbf{X}, \Theta)$ can directly estimate the ITEs using the diagnostic signals. We discard the initial models $f$ and $g$ during inference, making the process efficient and scalable. In our study, we use multi-layer perceptrons for $f$ and $g$, and a causal forest \cite{wager2018estimation, oprescu2019orthogonal} for $\theta$, which will be detailed next.

\subsection{Causal Forest\label{sec:cf}}
The causal forest is an extension of the random forest model \cite{breiman2001random} tailored to estimating the causal effects. The causal forest is an advanced version of the random forest model, specifically designed to estimate causal effects. While a random forest partitions data based only on input variables, a causal forest also considers the treatment variable ($A$) when making these partitions.

In a causal forest, each tree is split based on both the input variables ($\mathbf{X}$) and the treatment variable ($A$). This creates subgroups within each leaf of the tree, where the distribution of $A$ is balanced using propensity scores. Propensity scores \cite{joffe1999invited, rubin1996matching} help ensure that the treatment and control groups are comparable, effectively controlling for confounders. By balancing the treatment variable within each subgroup, the causal forest can more accurately estimate the treatment effect of $A$ on the outcome ($Y$). This method integrates causal inference principles with machine learning to capture complex relationships in the data while reducing bias. 

One of the key features of the causal forest is its ability to provide confidence intervals for its predictions. This is achieved using a technique called Bootstrap-of-Little-Bags \cite{diciccio1996bootstrap, athey2019generalized}. Confidence intervals are crucial when dealing with incomplete diagnostic information, as they give a measure of the uncertainty in the predictions. This feature helps address \textbf{Challenge 2} by providing more reliable decision-making under uncertainty.

We combine the causal forest with DML to create a more accurate and robust model for estimating ITEs. We call this combined approach \textbf{\model}. Compared to other ITE estimators like the meta-learner \cite{kunzel2019metalearners}, \name employs \model, which uses two-stage estimation and cross-fitting to reduce bias from model regularization and overfitting. Additionally, by incorporating causal forests, \model provides confidence intervals for its predictions, aiding in decision-making under uncertainty. More details on this will be provided in Sec.~\ref{sec:insurance}.

% \vspace*{-0.5em}
\section{\name Design and Implementation}
The foundation of \name rests on the core task of predicting ITE using the \model described in Sec.~\ref{sec:causal}. Various components have been devised to guarantee the secure and dependable deployment of the \model. We provide an overview of the overall architecture of \name in Sec.~\ref{sec:nutshell}, and delve into the details next.

\begin{figure}[t]
\centering
% \vspace*{-1.em}
\includegraphics[width=1\columnwidth]{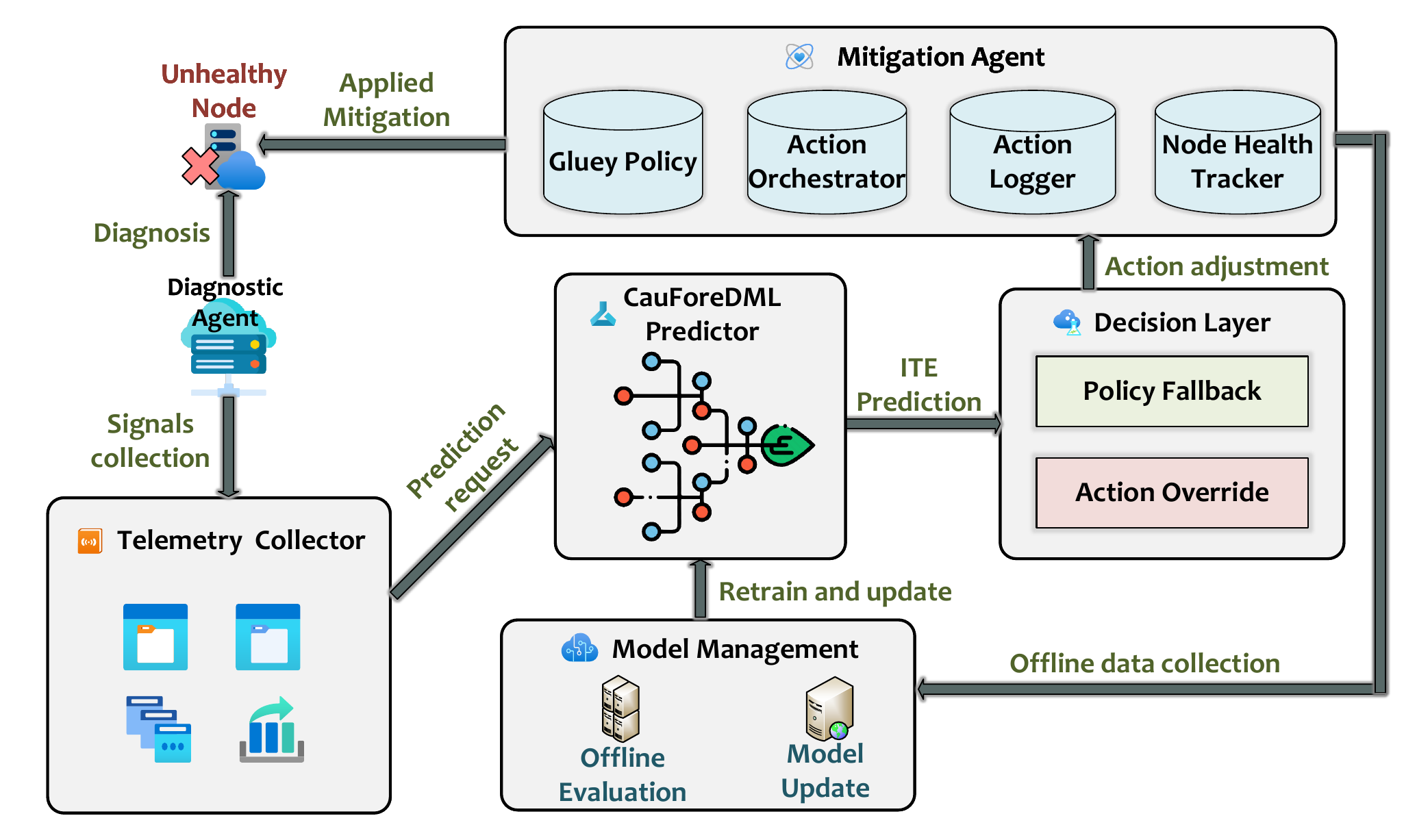}
\vspace*{-1.em}
\caption{The overall \name system architecture.
\label{fig:deoxys}}
% \vspace*{-1em}
\end{figure}

% \vspace*{-1em}
\subsection{System in a Nutshell\label{sec:nutshell}}
The proposed \name architecture is illustrated in Fig.~\ref{fig:deoxys}, and it comprises five distinct components that collectively support the mitigation selection policy. These components include a telemetry collector, a \model predictor, a decision layer, a model management module, and a mitigation engine. To recap the process depicted in Fig.~\ref{fig:unhealthy}, when an unhealthy event occurs, a diagnostic request is generated and sent to an agent, which then collects diagnostic signals and forwards them to the telemetry collector. If the unhealthy event exceeds a timeout threshold, \name is activated, and the \model (Sec.~\ref{sec:causal}) predicts the ITE between different mitigation actions. The predicted ITE values are then used as input to the decision layer, which combines them and adjust its predictions using policy fallback (Sec.~\ref{sec:insurance}) and action override mechanisms (Sec.~\ref{sec:counter-action}). The selected mitigation action is passed to the mitigation agent (Sec.~\ref{sec:agent}), to applies the action to the unhealthy node and logs and tracks the outcome of the request. The model management module controls the model update process (Sec.~\ref{sec:update}), tracking the online performance and updating the \model as needed. 
% We refer the detailed description of \model in Sec.~\ref{sec:causal}.
% and introduce the design of policy fallback and action override mechanisms in the decision layer and mitigation agent in the subsequent subsections.

% \vspace*{-1em}
\subsection{Policy Fallback \label{sec:insurance}}
In certain scenarios, the \model model may generate predictions with low confidence scores, indicating that the available diagnostic signals may not be sufficient for the model to make a confident recommendation. This situation may arise when the input data is incomplete or ambiguous, compromising the model's ability to accurately predict the outcome. The confidence scores are determined by the gap between the lower bound and upper bound of the prediction, with a larger gap indicating lower confidence. To address this issue, we have designed a fallback mechanism to the legacy mitigation policy in cases where the predicted ITE exhibits significant variance, \ie $\tau_u=\tau_{upper}- \tau_{lower} \geq \varepsilon$, and the predicted ITE is close to zero, \ie $|\tau| \leq \xi$, where $\varepsilon$ is the confidence threshold and $\xi \approx 0$. This indicates that the model is uncertain about its prediction and which action will lead to shorter downtime. In such cases, the legacy handcrafted policy can provide better performance and greater explainability. The fallback mechanism enhances the reliability of decision-making by incorporating the legacy policy when the data-driven model's predictions may be less reliable, thereby serving as a safeguard to ensure system reliability, addressing \textbf{Challenge 2} in Sec.~\ref{sec:challenge}.

% \vspace*{-0.5em}
\subsection{Action Override\label{sec:counter-action}}
In addition, predicted ITE from the \model may provide insights primarily focused on optimizing VM downtime, potentially overlooking other critical aspects such as node resource requirements and repetitive failures. Relying solely on this for decision-making may result in suboptimal outcomes. To mitigate this, we design two action override mechanisms, which allow for reverting decisions made by the \model to improve system reliability. \rv{These mechanisms also enable the \name to integrate human expertise into the decision-making process, providing a safeguard against potential errors made by the \model.}

\noindent\textbf{Changing \rd to \rb.}
The decision of whether to perform a \rb or a \rd action may also depends on the current capacity of the cloud system.  Recall that the \rd action requires the migration of all VMs to different nodes, which may be constrained by scarce node resources in the system. In such cases, the controller may opt for a \rb action to save more resource. To address this issue, we propose an action override rule in \name that overrides the \rd action before querying the capacity of the system, based on the quantity of the estimated ITE $\tau_i$. The absolute value of $\tau_i$ represents the potential downtime savings if one action is taken over the other. We set a override threshold for \rd as $\varpi$, such that if $\tau_i < 0$ and $|\tau_i| < \varpi$, we can override the \rd action with a \rb, resulting in the saving of more nodes required by \rd, while only increasing subtle downtime duration. \rv{The parameter 
$\varpi$ is set by minimizing a cost function that weighs the trade-off between downtime and resource usage, incorporating factors like revenue and user experience. \footnote{Due to the anonymization policies, we cannot disclose the specific formula of the cost function.} }

By leveraging this mechanism, \name is able to achieve a more favorable tradeoff between VM downtime and node resource, addressing the \textbf{Challenge 3} outlined in Sec.~\ref{sec:challenge}.

\noindent\textbf{Changing \rb to \rd.} 
In addition, given the data-driven nature of \name, it is susceptible to making errors, which can potentially result in undesirable system behaviors and a negative experience for consumers. For instance, in cases where hardware failures are the root cause of an unhealthy event, following \rb recommendations may not necessarily resolve the underlying issue. Persistently recommending \rb in such situations may lead to repeated VM interruptions within a short timeframe, as the underlying problem remains unresolved.

To address this challenge, another action override mechanism has been implemented in \name. This mechanism involves flagging a node as having an issue if unhealthy events occur repeatedly on the same node within a short period of time. In such cases, a \rd action may be enforced, regardless of \name's recommendation, and the node is marked as unallocatable for further investigation by human operators. This proactive approach helps to reduce the occurrence of unhealthy events and improves customer satisfaction, thus significantly enhancing the overall reliability of the system. This reduces the risk when \name makes a mistake, as discussed in \textbf{Challenge 4} in Sec.~\ref{sec:challenge}.

% \vspace*{-0.5em}
\subsection{Mitigation Agent\label{sec:agent}}
After receiving the action adjusted in the decision layer,  \name employs several components that work together as a mitigation agent to repair unhealthy nodes. These components are the Sticky Policy, Action Orchestrator, Action Logger, and Node Health Tracker, which communicate with each other and external services.

\begin{figure}[t]
\centering
% \vspace*{-1.em}
\includegraphics[width=1\columnwidth]{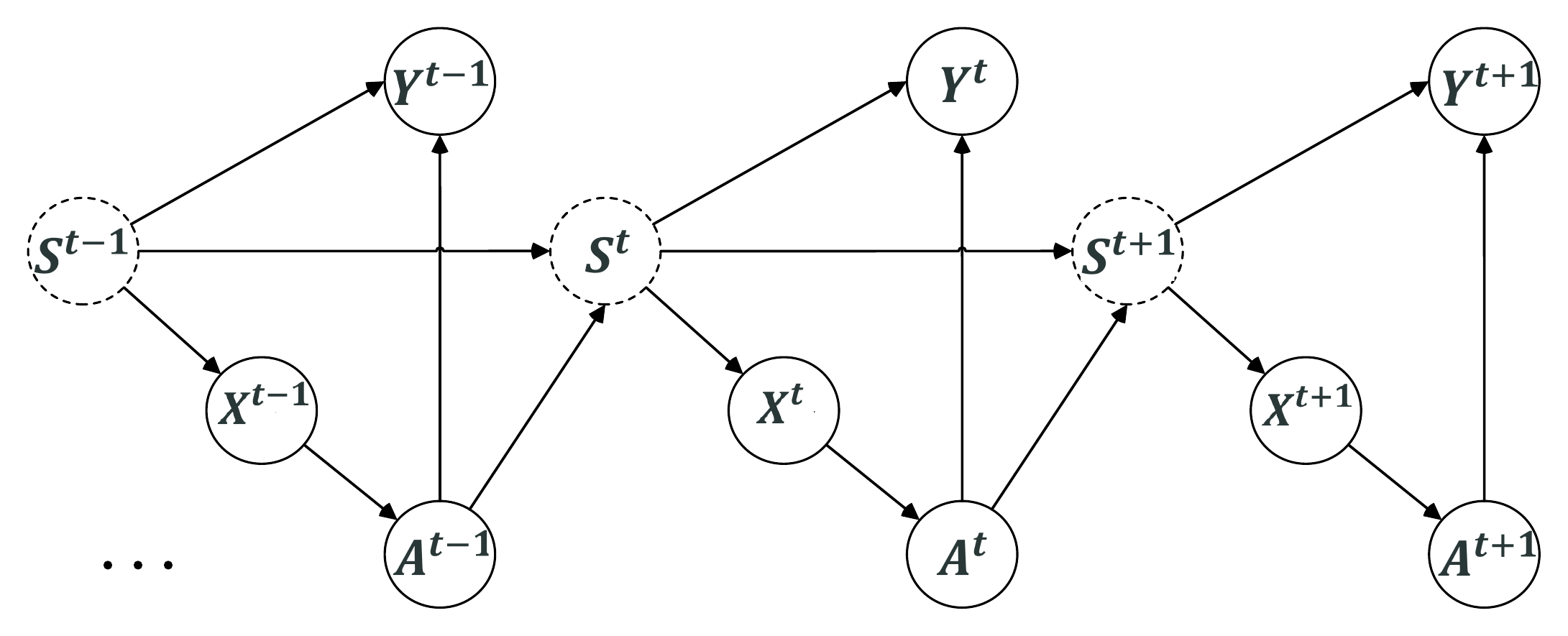}
\vspace*{-1.5em}
\caption{The dynamic causal graph of an unhealthy event.
\label{fig:dynamic_graph}}
% \vspace*{-1.3em}
\end{figure}

\noindent \textbf{Sticky Policy \cite{miorandi2019sticky}} The efficacy of \name is evaluated through \emph{policy-level} A/B testing compared against the legacy policy. This is opposed to the \emph{action-level} A/B testing between mitigation actions. Assigning a policy to a node or unhealthy request is a critical aspect of A/B testing. To ensure consistency throughout the testing process, it is necessary to assign a policy that remains the same for the duration of the experiment. In the classic A/B testing setting, units are assigned randomly under the assumption that they are independent and identically distributed (i.i.d). However, this assumption can be problematic when dealing with nodes in a cloud system, as the status of a node at time $t$, denoted by $S^t$, is not necessarily independent of previous behaviors or mitigation actions. Consecutive unhealthy events on the same node can be highly influenced by previous node statuses and behaviors. For example, if a node experiences hardware failures, a simple \rb action may not resolve the issue, resulting in new unhealthy events occurring in quick succession. If different policies are assigned to the same node during an experiment, the i.i.d assumption can be violated. We show this from a causal view in Fig.~\ref{fig:dynamic_graph}. The new time-dependent causal links between historical node statuses $S^{t-1}$, mitigation actions $A^{t-1}$, and the current node status $S^t$, create a Markovian process \cite{falmagne1988markovian, zhang2020microscope, zhang2021cloudlstm, chen2023imdiffusion} and violate the i.i.d assumption. 

To address this problem, we propose the use of a ``sticky policy''. For each node, the group is determined by hashing the node ID and policy group name. If a node is assigned to policy $P$ for an experiment, it will continue to use policy $P$ for subsequent unhealthy requests. This ensures that the policy remains consistent for each node throughout the A/B testing process, and the presence of time-dependent causation does not violate the backdoor criterion, as $\mathbf{X}^t$ continues to effectively block all backdoor paths originating from $A^t$ and leading to $Y^t$ in Fig.~\ref{fig:dynamic_graph}. This ensures the continued validity and reliability of our causal modeling approach.

\noindent \textbf{Action Orchestrator} The action orchestrator plays a critical role in executing the action plan developed during the policy tree walk session. It follows instructions from the decision layer and sticky policy, making API calls to the relevant compute managers to implement actions. As actions can be implemented by different managers, the orchestrator executes them asynchronously to prevent blocking.

\noindent \textbf{Action Logger} Effective logging is important for data analysis, offline data collection, and evaluating different mitigation policies. In \name, the logging format must record not only the chosen action but also the associated predicted ITE and corresponding confidence intervals. The mitigation agent logs the unhealthy timestamp, action timestamp, experiment name, model type, model name, model version, predicted ITE, confidence interval, chosen action, chosen action parameters, any triggers for action override or policy fallback, and the reasons for those triggers.

\noindent \textbf{Node Health Tracker} The node health tracker is responsible for monitoring node and VM health during the mitigation process. It tracks whether a VM is interrupted during mitigation and the time of its recovery. Additionally, it detects unsolvable node failures, such as hardware issues, and marks the node as unallocatable instead of waiting for resource brokers to allocate resources.

\subsection{Online Model Update\label{sec:update}}
The cloud environment is dynamic and continuously evolving due to factors like system updates and changes in customer workloads. To keep the \name system effective, we have designed a model update process involving several key steps. Initially, we collect recent unhealthy events from the past month to update the training dataset, which is used to retrain the DML model at the core of our system. This update ensures that the model incorporates the latest data patterns and trends.

Once the model is retrained, we conduct offline evaluations to assess its performance. This evaluation involves testing the model on a newer dataset that was not used during training, ensuring its ability to generalize effectively to new data. Subsequently, upon confirming that the new model outperforms the existing one, we deploy it online, replacing the old model. This process occurs on a regular weekly basis.

By adhering to this process, the \name system consistently utilizes the most accurate and up-to-date model to determine the optimal mitigation action for each unhealthy event. This proactive approach to model maintenance addresses data drift and minimizes the risk of using outdated models, thereby maintaining high performance levels over time.

\section{Offline Evaluation}

The \model is trained offline using a dataset from the real production environment. The model is constructed using the widely recognized causal inference tools \texttt{dowhy} \cite{dowhypaper} and \texttt{EconML} \cite{econml}. 

\subsection{Compared Methods}

Given that the ground-truth optimal mitigation action is unknown, we have devised a simulator to emulate the behaviors of unhealthy events within the cloud infrastructure at \company. This simulator serves as a tool to assess the effectiveness of \name and enables offline testing of various mitigation policies. It is constructed based on historical unhealthy events, encompassing various types of failures that have been verified by engineers to maximize fidelity. For comparison, we assess the performance of \emph{(i)} a \textbf{random} algorithm, \emph{(ii)} the \textbf{legacy policy} currently deployed online, \emph{(iii)} \textbf{Narya} \cite{levy2020predictive}, which employs bandit algorithms to explore and optimize mitigation policies in online environments, and \emph{(iv)} \textbf{Nenya} \cite{wang2022nenya}, which utilizes hierarchical reinforcement learning to develop an optimal mitigation policy in real-time. Both Narya and Nenya necessitate learning within an online production environment, which includes a cold-start period \cite{beggs2005convergence, zhang2022quickskill} before reaching convergence. Within the scope of causal inference methods, we also compare the accuracy of estimation among various approaches, as detailed in  Sec.~\ref{sec:causal_compare}.

\subsection{Offline Dataset Analysis\label{app:analysis}}

\begin{figure}[t]
\centering
% \vspace*{-1.em}
\includegraphics[width=1\columnwidth]{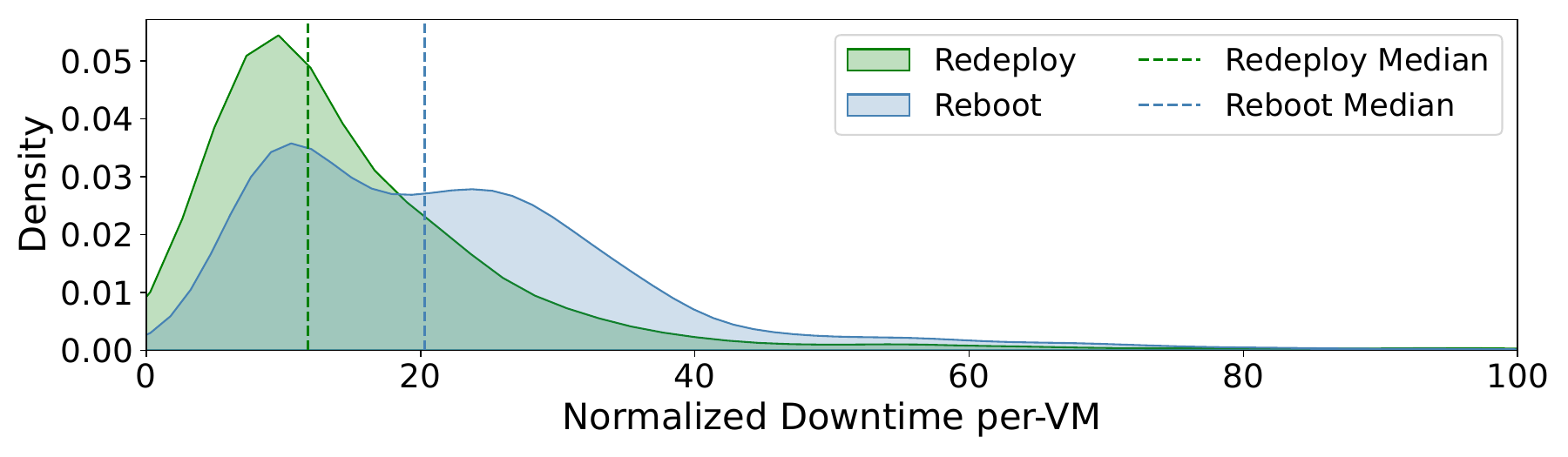}
\vspace*{-1.em}
\caption{The normalized average VM down time distributions for both \rb and \rd in the offline dataset. \label{fig:mitigation_distribution}}
% \vspace*{-1.em}
\end{figure}
We collected an offline dataset from the production environment spanning a duration of one month to train the initial \model model. This dataset comprises a total of 20,218 unhealthy events, with each sample containing diagnostic signals $\mathbf{X}$, the mitigation action $A$, and the average VM downtime $Y$. The data is generated based on the legacy policy and is purely observational in nature.

To gain insights from the offline dataset that used to train the causal model, we conducted a brief analysis. In Fig.~\ref{fig:mitigation_distribution}, we present the distributions of average VM downtime (\ie $P(Y|A=0)$ and $P(Y|A=1)$) as well as their median values for both \rb and \rd. Note that due to the company policy, we hide the actual time and normalize as time unit. Observe that in general, opting for \rd results in shorter VM downtime compared to \rb. However, this is subject to the bias discussed in Sec.~\ref{sec:backdoor}, and consistently selecting the \rd action may not necessarily be optimal. Moreover, the substantial overlap between $P(Y|A=0)$ and $P(Y|A=1)$ signifies the existence of a ``gray area'' and a blend in VM downtime concerning the mitigation action. 

\rv{
We also observe an intriguing phenomenon where nodes undergoing the \rb action exhibit a bi-modal distribution in downtime. This can be attributed to differing recovery outcomes post-\rb. In the first mode, some VMs recover rapidly, resulting in minimal downtime. In contrast, the second mode occurs when certain VMs fail to recover immediately, necessitating additional mitigation steps, which prolongs the downtime. This bi-modal distribution reflects the variability in reboot outcomes and recovery processes.
}

\subsection{Offline Performance Comparison}
\begin{table}[t]
\centering
\caption{The offline comparison of mitigation policy in 10,000 simulated unhealthy events.\label{tab:offline}}
% \vspace*{-1em}
\resizebox{\columnwidth}{!}{
\begin{tabular}{l|ccccc}
\hline
Policy  & Random  & Nerya & Nenya & \textbf{\name} \\ \hline
Downtime & 12.1       & 6.4   & 5.6   & \textbf{4.8}   \\ \hline
Convergence Events  & 0   & 136 & 1,350 & 0  \\ \hline
\end{tabular}
}
\end{table}

In Table~\ref{tab:offline}, we present an offline performance comparison of various mitigation policies using 10,000 simulated unhealthy events. The time units have been normalized and anonymized for consistency. Our results indicate that \name achieves the lowest downtime of 4.8 units, which is  a 14.3\% improvement over the best baseline, Nenya. This positions \name as a more reliable mitigation solution for the online environment. In contrast, Narya, which employs bandit algorithms but disregards the diagnostic signals as contextual information, performs worse. These findings underscore the superiority of data-driven policies over handcrafted ones, as it is challenging to design rules that generalize effectively across all cases.

We also compare the number of unhealthy events required for convergence across all methods. Our proposed method, \name, is trained on an offline dataset, thereby necessitating no exploration in the online environment and consistently delivering robust performance from the outset. In contrast, Narya and Nenya require 136 and 1,350 events, respectively, to achieve convergence. This translates to days or even weeks of exploration before these methods can deliver stable performance if deployed online, potentially adversely affecting user experience. This underscores the advantage of \name, which eliminates the cold start by training offline.

\subsection{Comparison with Causal Baselines\label{sec:causal_compare}}
In addition, we compare several variations of DML models with the \model  employed in \name on the same offline test set. These baselines include \emph{(i)} \textbf{LinearDML}: This baseline uses an unregularized linear model as the final estimator \cite{chernozhukov2018double}. \emph{(ii)} \textbf{SparseLinearDML}: This baseline employs an $l1$-regularized linear model as the final estimator, aiming for sparsity \cite{buhlmann2011statistics}. \emph{(iii)} \textbf{KernelDML}: This baseline utilizes a kernel function to better capture non-linearity in the effect model \cite{nie2021quasi}. \emph{(iv)} \textbf{NonParamDML}: This baseline makes no assumptions on the form of the effect model and uses \texttt{XGBoost} \cite{chen2015xgboost} as the final estimator \cite{kennedy2022semiparametric}.

\begin{table}[t]
\centering
\caption{Performance evaluation on different causal models.\label{tab:score}}
\begin{tabular}{c|c}
\hline
\textbf{Model}               & \textbf{Loss $\psi$} \\ \hline
LinearDML           & 12.1        \\ \hline
SparseLinearDML     & 8.3         \\ \hline
KernelDML           & 6.4         \\ \hline
NonParamDML         & 5.6         \\ \hline
\textbf{\model}     & \textbf{4.8} \\ \hline
\end{tabular}
% \vspace*{-2em}
\end{table}

We utilize the error function $\psi$ averaged on the test set as a quantitative measure to assess the accuracy of ITE estimation on a test set, defined in Eq.(\ref{eq:loss}). A lower value of $\psi$ indicates higher accuracy in estimating ITE. The results are presented in Table~\ref{tab:score}. Notably, \model, integrated into \name, achieves the best performance, demonstrated by the lowest $\psi$ score. Compared to the other baselines, \model achieves up to a 15\% reduction in final-stage loss, underscoring the superiority of combining DML and causal forest for ITE estimation. This highlights \model's effectiveness as a precise and robust ITE estimator in real-world production environments.

\subsection{Offline Counterfactual Analysis\label{app:counterfactual}}
\begin{figure}[t]
\centering
% \vspace*{-1.em}
\includegraphics[width=\columnwidth]{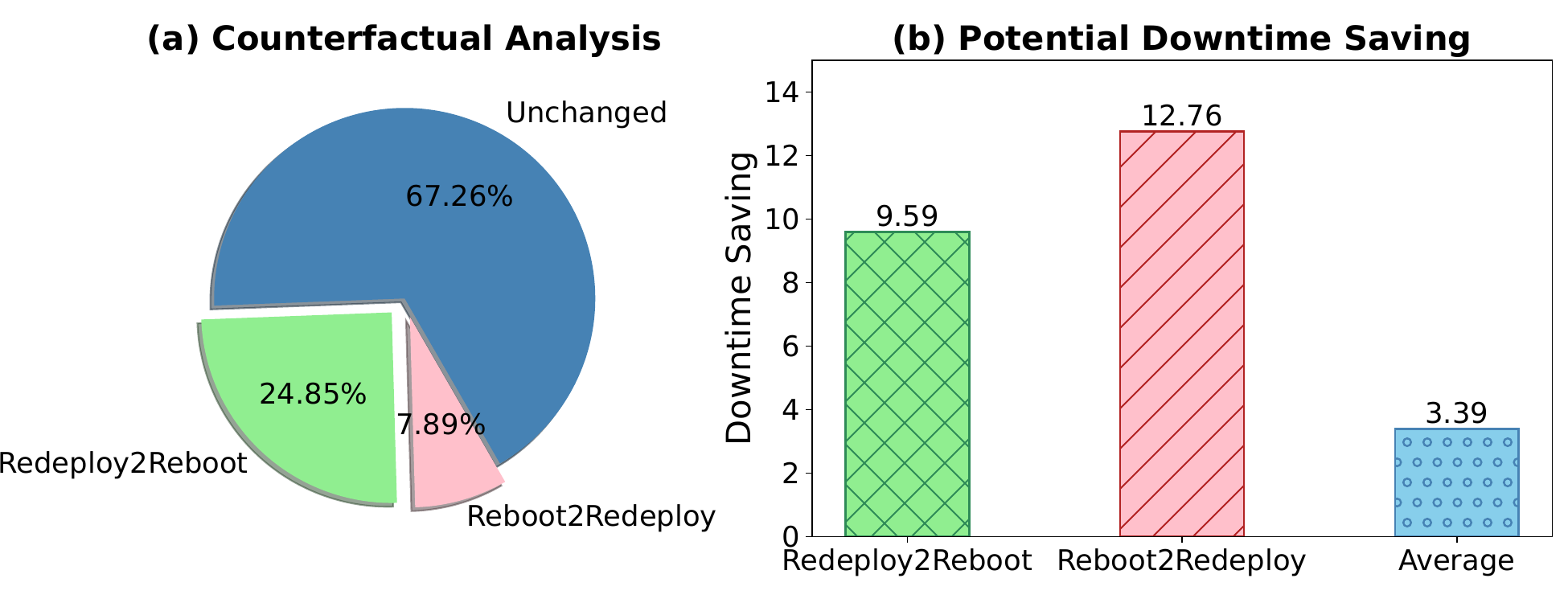}
\vspace*{-1.5em}
\caption{A counterfactual analysis on the offline dataset generated by the legacy policy. 
\label{fig:counterfactual}}
% \vspace*{-1.5em}
\end{figure}
The capability of causal inference empowers us to conduct a counterfactual analysis \cite{morgan2015counterfactuals} on the offline dataset using the \model. This analysis involves estimating the duration of downtime that would have occurred if a different action $A'$ had been taken to represent a counterfactual scenario, \ie $A' = 1-A$. Such a ``what-if'' analysis allows us to revisit the legacy policy in a postmortem manner and shed light on the potential downtime savings that could be achieved by transitioning to the causal policy employed in \name. \rv{It is important to note that these analysis results are based solely on offline data collected under the legacy policy and are not associated with the action override mechanisms discussed in Section~\ref{sec:counter-action}.}

Fig.~\ref{fig:counterfactual} presents the results of the counterfactual analysis conducted on the legacy policy, assuming a different action is taken. In subplot (a), we present the estimated proportion of unhealthy requests that could result in shorter downtime if a different action $A'$ had been taken. Notably, we observe that the action chosen by the legacy policy, which corresponds to 67.26\% of samples, aligns with the causal policy. However, the \model predicts that a different action should be taken for 33.74\% of unhealthy requests, resulting in reduced downtime compared to the original action. Interestingly, the \model estimator suggests that 24.85\% of the mitigations that were originally assigned to the \rd should switch to the \rb, which is more than three times the proportion of mitigations that should switch from the \rb to \rd. This underscores the potential benefit of leveraging causal inference in making policy decisions.

We further analyze the predicted downtime savings for each switching group. In the group where the \rd  should be taken instead of the \rb, the average downtime saving is estimated to be 9.59 units, whereas in the group where the \rb should be taken instead of the \rd , the average downtime saving is projected to be 12.76 units. This suggests a potential average reduction in downtime of 3.39 units if the legacy policy is replaced with the \model, and demonstrates the potential benefits of leveraging \model for guiding mitigation.

% \vspace*{-0.5em}
\section{Online Assessment in Production\label{sec:online}}
We have deployed \name as a core unhealthy mitigation engine in the cloud infrastructure of \company. To compare its performance against the legacy policy, we have conducted \emph{policy-level} A/B testing in production with comprehensive evaluations. Our A/B testing aims to answer the following research questions (RQs):
\begin{itemize}[leftmargin=*]
    \item \textbf{RQ1:} What is the effectiveness of \name in reducing VM downtime and affecting AIR?
    \item \textbf{RQ2:} What impact does \name have on other system metrics, such as blackout and capacity?
    \item \textbf{RQ3:} How can we leverage model uncertainty to design a reliable policy fallback?
    \item \textbf{RQ4:} Under what conditions should we override \rd with \rb, and what are its effects?
    \item \textbf{RQ5:} Under what conditions should we override \rb with \rd, and what are its effects?
\end{itemize}

% \vspace*{-0.5em}
\subsection{Deployment Scale and Experiment Method}
We conducted \emph{policy-level} A/B testing on 23 production regions in the cloud infrastructure of \company, spanning a duration of over 2 months, to compare the performance of \name and the legacy policy\footnote{The A/B testing is conducted to compare policies rather than actions, which is distinct from the action-level A/B testing introduced in Sec.~\ref{sec:intro} and~\ref{sec:backdoor}. Policy-level A/B testing is acceptable and low-cost to the infrastructure.}. This accounted for approximately one-third of the total nodes in production. Furthermore, for the sake of evaluation, we have also randomly designated two small sets of nodes to consistently apply either \rd or \rb actions. Due to their limited number, the impact on the overall cloud infrastructure is expected to be negligible. The experiment commenced on February 25\textsuperscript{th}, 2023, and ended on June 25\textsuperscript{th}, 2023. This coverage encompassed a significant portion of nodes in the cloud, as well as a substantial number of unhealthy requests. Nodes were randomly hashed into either the \name group or the legacy policy group with equal probability using the sticky policy, as described in Sec.~\ref{sec:agent}. Note that only unhealthy requests that lacked deterministic signals for diagnosis and exceeded the unhealthy timeout were included in the A/B testing. This accounts for 55\% of the total unhealthy requests. We have opted not to compare our \name with the frameworks proposed in \cite{levy2020predictive} and \cite{wang2022nenya} in production, as they demand substantial enhancements to our existing infrastructure in order to facilitate the requisite real-time interaction.

\subsection{Downtime \& Interruption Reduction (RQ1)}

\begin{table}[t]
\caption{The improvement of AVD and AIR achieved by \name  compared to other policies in the A/B testing. \label{tab:eva}}
% \vspace*{-1em}
\resizebox{\columnwidth}{!}{
\begin{tabular}{cccccccc}
\hline
\multirow{2}{*}{Policy} & \multirow{2}{*}{Sample} & \multirow{2}{*}{\begin{tabular}[c]{@{}c@{}}VM\\ count\end{tabular}} & \multicolumn{4}{c}{AVD gain} & \multirow{2}{*}{\begin{tabular}[c]{@{}c@{}}Avg.\\ air gain\end{tabular}} \\ \cline{4-7}
                        &                                                                             &                                                                     & P50   & P75   & P90   & Avg. &                                                                          \\ \hline
Legacy                  & 845                                                                         & 5,138                                                               & 49\%  & 45\%  & 44\%  & 53\% & 49.5\%                                                                   \\
\rb                     & 98                                                                          & 522                                                                 & 65\%  & 62\%  & 61\%  & 63\% & 54.3\%                                                                   \\
\rd                     & 106                                                                         & 559                                                                 & 25\%  & 23\%  & 23\%  & 24\% & 12.6\%                                                                   \\
\name                   & 802                                                                         & 4,587                                                               & -     & -     & -     & -    & -                                                                        \\ \hline
\end{tabular}
}
% \vspace*{-1.5em}
\end{table}

The KPIs utilized to evaluate the effectiveness of \name are the average VM downtime (AVD) and annual interruption rate (AIR), as described in Sec.~\ref{sec:kpi}. Table~\ref{tab:eva} presents the improvement achieved by \name compared to the legacy policy in terms of percentiles, and average values of AVD and AIR. In total, 845 and 802 unhealthy events trigger the legacy policy and \name during the A/B testing, with 5,138 and 4,587 VM were impacted respectively. Furthermore, the policies of always applying \rb and \rd actions were triggered by a limited number of unhealthy events (98 and 106 cases). Due to the small quantity, these events are not expected to exert a significant adverse impact.

Notably, \name exhibits lower AVD values across all percentiles as well as the average value, demonstrating significant improvement in comparison to the legacy policy\footnote{We hide the actual numbers of all metrics and only report the relative improvement, due to \company's confidentiality requirement.}. Specifically, \name achieves reductions of 49\%, 45\%, and 44\% in AVD for P50, P75, and P90, and a remarkable 53\% reduction in average AVD. Moreover, although AIR is not a direct optimization objective, \name still outperforms the legacy policy by a substantial margin of 49.5\%. This highlights the hidden correlation between AVD and AIR, with \name effectively addressing both VM downtime and interruption rate. It is crucial to note that the legacy policy serves as a strong baseline, having been designed by experienced engineers and employed in the \company cloud infrastructure for years. Given the extensive scale of the cloud infrastructure and the frequent occurrence of unhealthy events, the significant gains achieved by \name positively impact a large number of customers. 

Likewise, in comparison to the consistently applied \rb and \rd policies, our \name consistently demonstrates superior performance across all metrics, including AVD and AIR. This underscores the importance of customizing mitigation strategies for specific unhealthy events, as a one-size-fits-all approach is suboptimal. Notably, while the \rd policy outperforms the legacy approach, it necessitates a significant allocation of additional node resources to accommodate the redeployed VMs, in contrast to rebooting a node, which retains the recovered VMs on the same nodes. This resource-intensive nature of \rd is the principal reason for its non-deployment in real-time online scenarios.

\subsection{Node Interruption Assessment (RQ2)}

\begin{table}[t]
\centering
\caption{Blackout duration improved by \name.\label{tab:blackout}}
% \vspace*{-1em}
% \resizebox{\columnwidth}{!}{
\begin{tabular}{cccccc}
\hline
Blackout Duration      & Avg.  & P50   & P75   & P90   & P99   \\ \hline
Improvement & 3.6\% & 5.4\% & 3.4\% & 0.5\% & 2.6\% \\ \hline
\end{tabular}
% \vspace*{-1em}
\end{table}

\begin{table}[t]
\centering
\caption{Unallocatable metrics improved by \name.\label{tab:unallocatable}}
% \vspace*{-1em}
\resizebox{\columnwidth}{!}{
\begin{tabular}{cccccc}
\hline
Unallocatable Metric      & Ratio  & Avg.  & P50  & P75    & P90     \\ \hline
Improvement & 0\%    & 12.0\%  & 0\%  & 12.8\% & 17.9\%    \\ \hline
\end{tabular}}
% \vspace*{-1em}
\end{table}

The mitigation policy introduces interruptions to unhealthy nodes, such as \emph{(i)} \textbf{Blackout}, which can cause blips and pauses to the deployed VM, leading to performance regressions such as jitters and slowdowns, and \emph{(ii)} temporally \textbf{Unallocatable}, which occurs during the duration and process of mitigation action and results in the node being out of capacity and unable to be allocated with new VMs.  Note that a shorter blackout/unallocatable duration and a smaller unallocatable rate indicate better system reliability. We evaluate these metrics to provide additional insight into the impact of \name, for a more comprehensive assessment.

We present the reduction improvement achieved by \name in terms of the blackout duration, unallocatable rate, and duration over the legacy policy in Tables~\ref{tab:blackout} and~\ref{tab:unallocatable}, respectively. Our results show that \name outperforms the legacy policy, with improvements in terms of average and all percentile values for blackout duration. This means that customers will experience less performance regression, resulting in a seamless user experience. Furthermore, as shown in Table~\ref{tab:unallocatable}, while both policies exhibit similar unallocatable rates, employing \name can shorten the average unallocatable duration by 12\%. This is crucial for the system's capacity, as a shorter unallocatable duration means that the overall cluster can accommodate more VMs in a given period.

% \vspace*{-0.5em}
\subsection{Policy Fallback  on Model Uncertainty (RQ3)}
% \vspace*{-1.em}
\begin{figure}[t]
\centering
% \vspace*{-1.em}
\includegraphics[width=1\columnwidth]{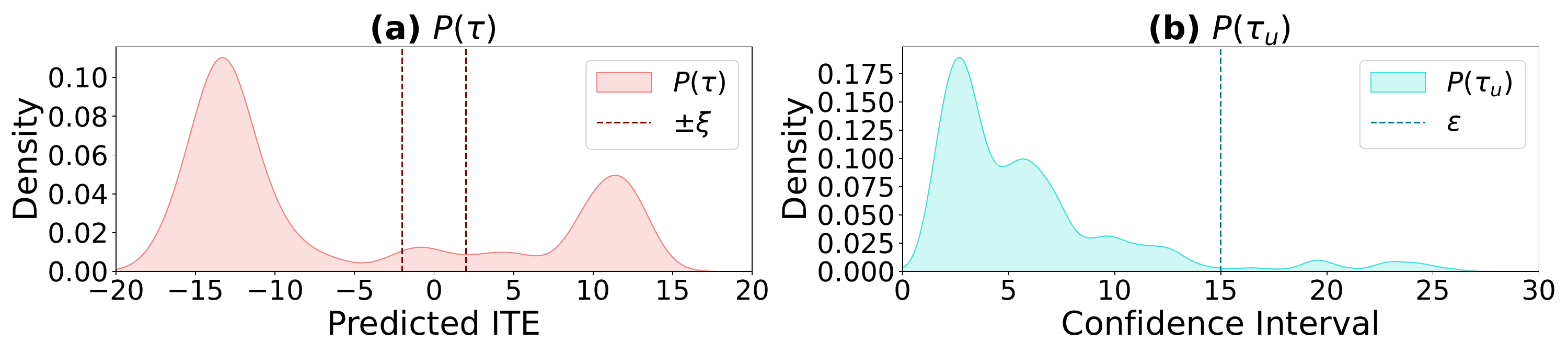}
\vspace*{-2.2em}
\caption{The distribution of the (a) estimated ITE $\tau$ and (b) its confidence interval $\tau_u$.
\label{fig:tau}}
% \vspace*{-1em}
\end{figure}
As discussed in Sec.~\ref{sec:insurance}, we have designed a callback mechanism based on the predicted ITE $\tau$ and its confidence interval $\tau_u$. The purpose of this fallback  mechanism is to enhance the reliability of the system in cases where the model feels uncertain or when the diagnostic signals are insufficient to make a decision. In Fig.~\ref{fig:tau}, we present the distribution of ITE estimation $\tau$ and its confidence interval $\tau_u$. We observe that the distribution of $P(\tau)$ exhibits a bimodal pattern, with only a small proportion of $\tau$ values being close to zero. Furthermore, the \name model tends to make confident predictions for the majority of samples, but occasionally its predictions remain hesitant. Based on the insights from Fig.~\ref{fig:tau}, we have set the threshold parameters $\xi = 1$ and $\varepsilon = 15$, such that if $|\tau| < 1$ and its $\tau_u > 15$, we consider the model to be not confident, and therefore rollback to the legacy policy to implement a safer mitigation strategy. In production, approximately 2.3\% of unhealthy requests trigger this fallback.

In a separate set of A/B testing, activating this fallback mechanism achieved similar VM downtime while reducing AIR by 18.1\% compared to the group without it. This demonstrates its effectiveness in improving system reliability. It also validates that handcrafted policies are effective when the model is uncertain.

% \vspace*{-0.5em}
\subsection{Overriding \rd with \rb (RQ4)}
\begin{figure}[t]
\centering
% \vspace*{-1.em}
\includegraphics[width=1\columnwidth]{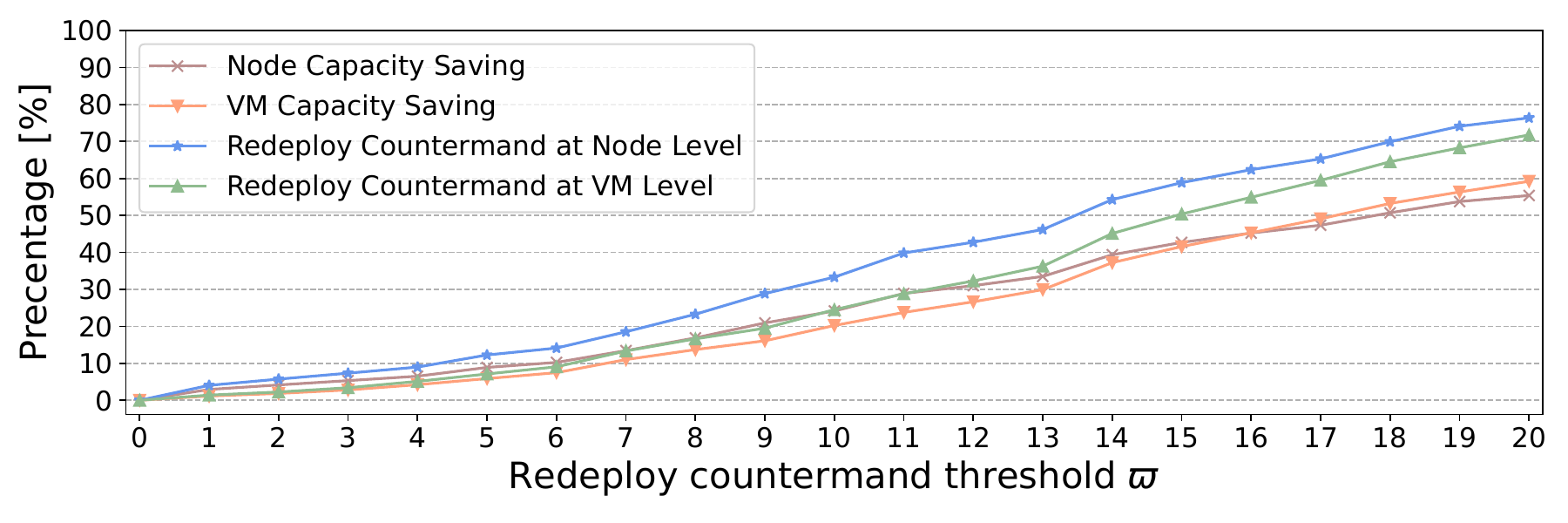}
\vspace*{-2em}
\caption{Potential node/VM saving as well as their override ratio w.r.t the \rd override threshold $\varpi$.
\label{fig:rd_countermand}}
% \vspace*{-1.5em}
\end{figure}
We conducted an evaluation of the first action override mechanism in Sec.~\ref{sec:counter-action}, where we choose to use \rb instead of \rd, even when the estimator predicts that taking \rd would result in shorter VM downtime. The aim of this mechanism is to save more node resources that would be required by \rd, with minimal impact on downtime. In Fig.~\ref{fig:rd_countermand}, we present the offline simulation results of the node and VM resource saving proportion, as well as their override ratio, within the requests that recommended the \rd action, with respect to the \rd override threshold $\varpi$. Recall that $\varpi$ represents the maximum predicted downtime saving by taking \rd, when we override the \rd action in exchange for node resource. The figure indicates that by setting $\varpi = 1$, we can exchange 3\% of the nodes requesting \rd while sacrificing only 1 unit of VM downtime. This results in substantial resource savings and enhances cluster capacity, presenting a favorable tradeoff between VM downtime and node resources. We have set this value in production for system capacity consideration.

% \vspace*{-0.5em}
\subsection{Overriding \rb with \rd (RQ5)}
Additionally, we have implemented a feature called \texttt{RepeatCnt} to keep track of the number of unhealthy events that have occurred on the same node within the past 10 days. If the \texttt{RepeatCnt} exceeds a certain threshold, it indicates that the node has experienced repetitive failures that cannot be easily resolved by using \rb alone. In such cases, we override the model's decision and opt for \rd regardless of its recommendation. Furthermore, we mark the node as unallocatable and initiate a human investigation to identify the root cause of the repetitive failures. In our production environment, we have empirically set the threshold for triggering this override mechanism at \texttt{RepeatCnt} $>10$. As a result, approximately 8\% of the unhealthy requests trigger this mechanism, out of which 13\% were originally recommended with \rb but were overridden in favor of \rd. Upon post-analysis of these cases, \emph{it was observed that more than 96\% of these nodes were susceptible to uncorrectable errors if the \rb action was taken.} Our design effectively overrode this to \rd, successfully correcting the mitigation action. This overriding mechanism ensures that nodes experiencing recurrent failures receive the necessary mitigation measures, even when the model's prediction recommends a different course of action.

% \subsection{Case Study}
% \vspace*{-0.5em}
\section{Policy Interpreter\label{sec:interpreter}}
\begin{figure}[t]
\centering
% \vspace*{-1.em}
\includegraphics[width=1\columnwidth]{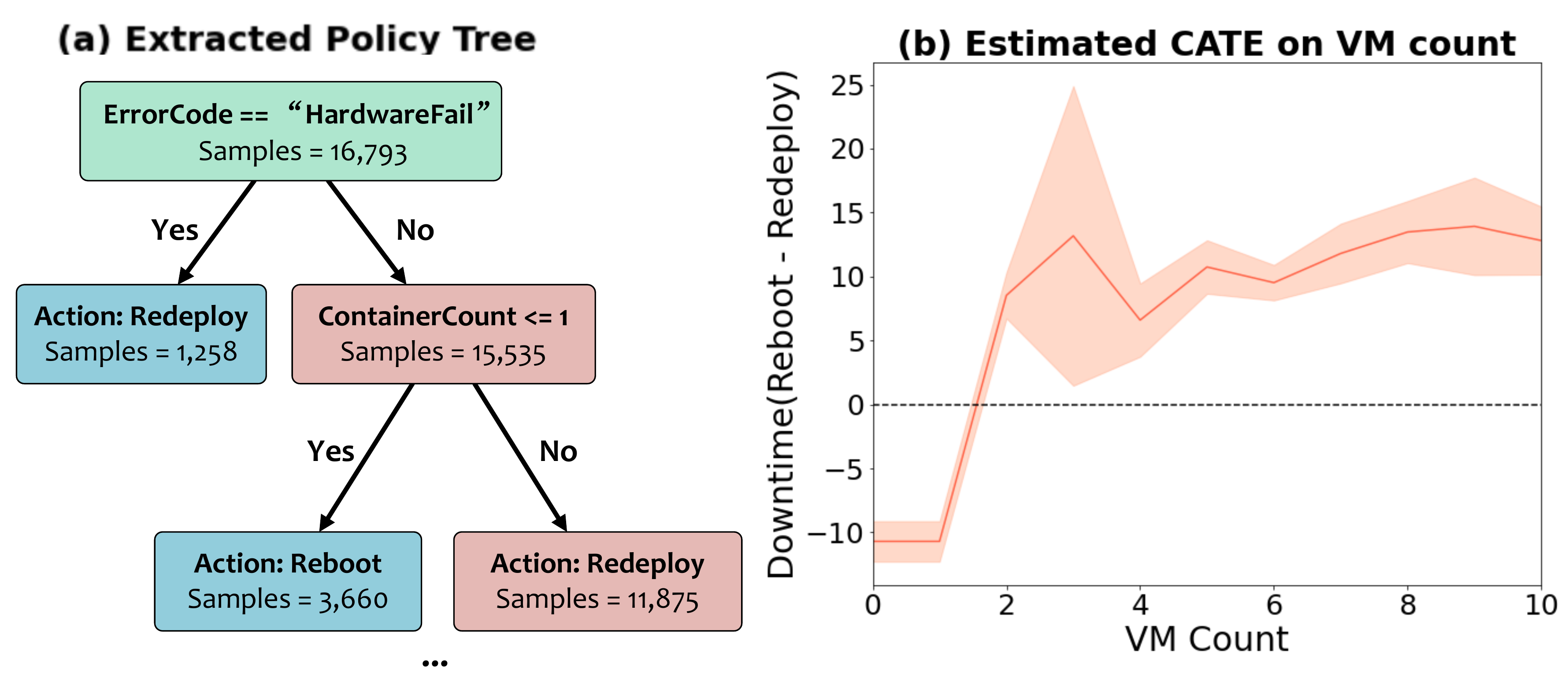}
\vspace*{-2em}
\caption{Policy interpreter. (a) The partial logical policy tree extracted; (b) The CATE conditioned on VM count.
\label{fig:interpreter}}
% \vspace*{-1.5em}
\end{figure}
Finally, to enhance interpretability, the \texttt{EconML} library \cite{econml} provides a \texttt{SingleTreePolicyInterpreter} tool that offers a summary of the key diagnostic signals that explain the most significant differences in responsiveness to a mitigation action. This tool trains a shallow decision tree \cite{safavian1991survey} using the treatment effect obtained from the \model on a small set of diagnostic signals. The decision tree learns an ``if-else'' like policy by splitting on cutoff points that maximize the treatment effect difference in each leaf. The resulting logical policy provides insights into why the \model makes a particular prediction, enhancing the interpretability and unveils the system behaviors in unhealthy events.

The initial two layers of the deduced logical policy are displayed in Fig~\ref{fig:interpreter} (a), with the subsequent layers omitted for brevity. Notably, the interpreter discloses that the foremost pivotal diagnostic signal driving the decision-making process for mitigation pertains to instances where the node reports a hardware failure error code. This rationale aligns with the inherent logic that, in the presence of hardware failures, resorting to a \rb is unlikely to rectify the issue, thereby necessitating the adoption of the \rd action. In addition, the interpreter highlights the significance of the count of live VMs on the unhealthy node. When the VM count is less than or equal to 1, the recommended action is to take \rb. On the other hand, when the VM count is greater than 1, the recommended action is to take \rd.

Based on the above findings, we further investigate the correlation of the treatment effect of the mitigations, conditioned on the alive VM count, as estimated by the Conditional Average Treatment Effect (CATE), as shown in Fig.~\ref{fig:interpreter} (b). The x-axis of the figure represents the VM count, and the y-axis represents the estimated downtime difference if taking \rb instead of \rd. Interestingly, the CATE suggests that we should opt for \rd instead of \rb with an increase in the VM count, as the CATE increases. It shows negative values at VM count$=1$. This is aligned with the pivot point identified by the interpreter.

These findings prompt us to delve deeper into the their underlying factors. In general, a \rb action tends to result in shorter downtime when the unhealthy signal is a false alarm, which occur frequently in cloud environments. However, there is a risk that the downtime will become significantly longer if the node indeed has unsolvable problems. On the other hand, the \rd action is considered safer as it relocates all VMs to a different node. Nonetheless, \rd increases the VM downtime when the node is actually healthy, as \rb is faster in such cases. Conditioning the mitigation action on the number of VMs on the node allows for a tradeoff between risk and downtime. If there are many VMs on the node, it may be advisable to avoid the risk of \rb, as the cost of a failed \rb could be high. However, if there are only very few VMs, it may be beneficial to exploit the potential benefits of \rb. These findings shed light on an interesting system tradeoff, and it is remarkable that this is learned by \name.

% \vspace*{-0.5em}
\section{Lessons learned and Limitations}

While \name primarily focuses on determining ``what to take'' as a mitigation action when unhealthy events occur, it is important to note that some of these events may be false positives, such as when the loss of heartbeat signal is caused by network issues rather than actual failures. In such cases, the node may not be truly unhealthy and may come back ready soon without any action needed. Mitigating these false positive events can lead to unnecessary VM interruptions and result in higher AIR. Therefore, developing a model to determine ``when to take'' action, \ie waiting for a certain duration before mitigation, can be a valuable addition to the framework. This approach can help improve both AVD and AIR, as unnecessary mitigation takes time and increases VM interruptions. We acknowledge that this is a potential area for future research.

We also recognize that, while historical unhealthy events, node status snapshots, and actions taken are crucial for understanding the system behavior.  The current \name does not consider this sequential information yet. This may limit the performance of the model. The next step is to build an online database to buffer such historical behaviors, allowing the model to leverage this information for more accurate action recommendations. This can potentially enhance the overall performance and effectiveness of the \name.

% \vspace*{-0.5em}
\section{Related Work}
% In this section, we review related research and industry practice on failure mitigation and causal inference.
\noindent \textbf{Failure Mitigation} plays a crucial role in modern cloud infrastructure \cite{li2020gandalf}, as it significantly impacts customers' experience \cite{patterson2002recovery, zhang2018deepview}. Several existing approaches have been proposed to address this challenge. Narya \cite{levy2020predictive} is a proactive mitigation engine in Azure that utilizes multi-arm bandit techniques \cite{slivkins2019introduction} to explore optimized mitigation policies in online environments. It has been further improved by NENYA \cite{wang2022nenya}, which integrates failure prediction \cite{lin2018predicting, ma2022empirical, liu2022multi} and mitigation into an end-to-end framework. Narya is considered as a non-contextual approach, which fails to model system heterogeneity given different node status.
Moreover, these alternative methodologies necessitate online interactions with the live production environment for model updates, which in turn entail cold-start periods. Such operational requirements have the potential to induce performance regressions within the system. In contrast, \name adopts an offline training approach, eschews frequent updates, and consistently yields dependable results. Consequently, this approach stands as a more favorable choice for production scenarios.

\noindent \textbf{Causal Machine Learning}
Causal machine learning has emerged as a prominent topic in recent research due to its ability to automatically discover and estimate complex causal effects in high-dimensional data, even in the presence of unobserved confounders \cite{kaddour2022causal, zhang2019deep, athey2015machine}. One of the key applications of causal machine learning is Individual Treatment Effect (ITE) estimation, which is also utilized in our proposed  \name \cite{zhang2021unified}. ITE estimation has been widely employed in various domains such as recommender systems \cite{wang2021deconfounded, chen2023bias} and healthcare applications \cite{sanchez2022causal, prosperi2020causal}. Commonly used methods for ITE estimation, such as S-learner \cite{xu2022learning}, T-learner \cite{olaya2020uplift} and X-learner \cite{zhao2019uplift}, are popular tools in the field. However, these methods often lack the ability to provide confidence intervals for their predictions, which could indicate the uncertainty of the models. In contrast, our proposed \name approach employs DML \cite{chernozhukov2018double} and causal forest \cite{wager2018estimation} to  deliver confidence scores for policy fallback. This improves the reliability of data-driven methods applied in cloud systems.

% \vspace*{-1em}
\section{Conclusion}
In this paper, we introduce \name, a causal unhealthy node mitigation engine for large-scale cloud infrastructures. \name leverages observational data to learn the causation between different mitigation actions and VM downtime, allowing it to choose the most effective action to minimize downtime during unhealthy events. The reliability and trustworthiness of \name are ensured through the incorporation of policy fallback and action override mechanisms, improving the robustness of the system. Furthermore, it has the capability to extract logical rules from data, elucidating its decision-making process and enhancing system interpretability.
Notably, we deployed \name as a key unhealthy node mitigation engine in a large-scale cloud infrastructure at \company. Our observations have shown that, on average, \name reduces VM downtime by \saving compared to a legacy mitigation policy, while leading \airsaving lower VM interruption rate. This demonstrates the effectiveness of \name in enhancing the performance and stability of the cloud computing platform, 
thereby providing better customer experience and potentially leading to huge business value.

% \newpage
\bibliographystyle{ACM-Reference-Format}
\balance
\bibliography{sample}

\end{document}